\documentclass[fleqn,usenatbib]{mnras}

\usepackage{newtxtext,newtxmath}

\usepackage[T1]{fontenc}

\usepackage{graphicx}
\usepackage{amsmath}

\usepackage{amssymb}
\usepackage[utf8]{inputenc}
\usepackage{graphics,epsfig,ulem,times}
\usepackage{pgfplots}
\usepackage[caption=false]{subfig}
\usepackage{xcolor}
\usepackage{xtab}
\usepackage{pifont}

\usepackage{array}
\newcolumntype{C}[1]{>{\centering\arraybackslash}p{#1}}
\newcolumntype{?}{!{\vrule width 2.5\arrayrulewidth}}
\usepackage{multirow, makecell}
\usepackage[flushleft]{threeparttable}
\usepackage{caption}
\DeclareCaptionFormat{cont}{#1 (cont.)#2#3\par}

\pgfplotsset{compat=1.18}

\title[Disc-like kinematics in DSFGs]{KAOSS: turbulent, but disc-like kinematics in dust-obscured star-forming galaxies at $z$\,$\sim$\,1.3--2.6}

\author[J. Birkin et al.]{
Jack E.\ Birkin,$^{1,2,3}$\thanks{E-mail: jbirkin@tamu.edu}
A.\,Puglisi,$^{4,1}$\thanks{Anniversary Fellow}
A.\,M.\ Swinbank,$^{1}$
Ian Smail,$^{1}$
Fang Xia An,$^{5,6}$
S.\,C.\ Chapman,$^{7}$\newauthor
Chian-Chou Chen,$^{8}$
C.\,J.\,Conselice,$^{9}$
U.\,Dudzevi{\v{c}}i{\={u}}t{\.{e}},$^{10}$
D.\,Farrah,$^{11,12}$
B.\,Gullberg,$^{13,14}$
Y.\,Matsuda,$^{15,16,17}$\newauthor
E.\,Schinnerer,$^{10}$
D.\,Scott,$^{18}$
J.\,L.\ Wardlow,$^{19}$
and P.\ van der Werf$^{20}$
\\
$^{1}$Centre for Extragalactic Astronomy, Department of Physics, Durham University, South Road, Durham, DH1 3LE, UK\\
$^{2}$Department of Physics and Astronomy, Texas A\&M University, 4242
TAMU, College Station, TX 77843-4242, USA\\
$^{3}$George P. and Cynthia Woods Mitchell Institute for Fundamental Physics and Astronomy, Texas A\&M University, 4242
TAMU, College Station, TX 77843-4242, USA\\
$^{4}$School of Physics and Astronomy, University of Southampton, Highfield SO17 1BJ, UK\\
$^{5}$Purple Mountain Observatory, Chinese Academy of Sciences, 10 Yuanhua Road, Nanjing 210034, People's Republic of China\\
$^{6}$Inter-University Institute for Data Intensive Astronomy, University of the Western Cape, Robert Sobukwe Road, Bellville, Cape Town 7535, South Africa\\
$^{7}$Department of Physics and Atmospheric Science, Dalhousie University, Halifax, Halifax, NS B3H 3J5, Canada\\
$^{8}$Academia Sinica Institute of Astronomy and Astrophysics (ASIAA), No. 1, Section 4, Roosevelt Road, Taipei 10617, Taiwan\\
$^{9}$Jodrell Bank Centre for Astrophysics, University of Manchester, Oxford Road, Manchester UK\\
$^{10}$Max-Planck-Institut f\"{u}r Astronomie, K'onigstuhl 17, D-69117, Heidelberg, Germany\\
$^{11}$Department of Physics and Astronomy, University of Hawai'i, 2505 Correa Road, Honolulu, HI 96822, USA\\
$^{12}$Institute for Astronomy, 2680 Woodlawn Drive, University of Hawai'i, Honolulu, HI 96822, USA\\
$^{13}$Cosmic Dawn Center (DAWN), Denmark\\
$^{14}$DTU Space, Technical University of Denmark, Elektrovej 327, DK-2800 Kgs. Lyngby, Denmark\\
$^{15}$National Astronomical Observatory of Japan, Osawa 2-21-1, Mitaka, Tokyo 181-8588, Japan\\
$^{16}$Graduate University for Advanced Studies (SOKENDAI), Osawa 2-21-1, Mitaka, Tokyo 181-8588, Japan\\
$^{17}$Cahill Center for Astronomy and Astrophysics, California Institute of Technology, MS 249-17, Pasadena, CA 91125, USA\\
$^{18}$Department of Physics and Astronomy, University of British Columbia, 6224 Agricultural Road, Vancouver, BC V6T 1Z1, Canada\\
$^{19}$Department of Physics, Lancaster University, Lancaster, LA1 4YB, UK\\
$^{20}$Leiden Observatory, Leiden University, P.O. box 9513, NL-2300 RA Leiden, the Netherlands\\
}

\pubyear{2024}

\begin{document}

\setlength{\parskip}{0pt}
\label{firstpage}
\pagerange{\pageref{firstpage}--\pageref{lastpage}}
\maketitle

\begin{abstract}
We present spatially resolved kinematics of 27 ALMA-identified dust-obscured star-forming galaxies (DSFGs) at $z$\,$\sim$\,1.3--2.6, as traced by H$\alpha$ emission using VLT/KMOS near-infrared integral field spectroscopy from the ``KMOS-ALMA Observations of Submillimetre Sources'' (KAOSS) Large Programme. We derive H$\alpha$ rotation curves and velocity dispersion profiles for the DSFGs, and find that among the 27 sources with bright, spatially extended H$\alpha$ emission, 24 display evidence for disc-like kinematics. We measure a median inclination-corrected velocity at 2.2\,$R_{\rm d}$ of $v_{\rm rot}$\,$=$\,190\,$\pm$\,40\,km\,s$^{-1}$ and intrinsic velocity dispersion of $\sigma_0$\,$=$\,87\,$\pm$\,6\,km\,s$^{-1}$ for these {\it disc-like} sources. The kinematics yield median circular velocities of $v_{\rm circ}$\,$=$\,230\,$\pm$\,20\,km\,s$^{-1}$ and dynamical masses within 2$R_{\rm e}$ ($\sim$\,7\,kpc radius) of $M_{\rm dyn}$\,$=$\,(1.1\,$\pm$\,0.2)\,$\times$\,10$^{11}$\,M$_\odot$. Compared to less actively star-forming galaxies, KAOSS DSFGs are both faster rotating with higher intrinsic velocity dispersions, but have similar $v_{\rm rot}/\sigma_0$ ratios, median $v/\sigma_0$\,$=$\,2.5\,$\pm$\,0.5. We suggest that the kinematics of the DSFGs are primarily rotation supported but with a non-negligible contribution from pressure support, which may be driven by star formation or mergers/interactions. We estimate the normalisation of the stellar mass Tully-Fisher relation (sTFR) for the {\it disc-like} DSFGs and compare it with local studies, finding no evolution at fixed slope between $z$\,$\sim$\,2 and $z$\,$\sim$\,0. Finally, we show that the kinematic properties of the DSFG population are consistent with them evolving into massive early-type galaxies, the dominant $z$\,$\sim$\,0 population at these masses.
\end{abstract}




%
%
\section{Introduction}
\label{sec:intro}

Dust-obscured star-forming galaxies (DSFGs) at the peak of cosmic star formation ($z$\,$\sim$\,2) are massive and gas rich, with star-formation rates (SFRs) that are significantly higher than typical systems at this epoch \citep{tacconi06,magnelli12,bothwell13,swinbank14,miettinen17,dudzeviciute20,birkin21,shim22}. However, their kinematics are poorly understood due to a lack of spatially resolved observations. Are they predominantly turbulent merger-driven \citep[e.g.,][]{narayanan09,narayanan10,lagos20} systems, like the similarly infrared-bright local Ultra-Luminous Infrared Galaxy (ULIRG) population \citep[e.g.,][]{bellocchi16}? Or do they more closely resemble regular discs that are smoothly accreting gas from the intergalactic medium \citep[IGM;][]{keres05,dekel06,narayanan15,tacconi20}?

One of the most promising routes to test these competing theories is through integral field spectroscopy (IFS) in the rest-frame optical, which enables two-dimensional (2-D) mapping of the spatially resolved kinematics via nebular emission lines such as H$\alpha$ \citep[e.g.,][]{swinbank06a,forster-schreiber09,alaghband-zadeh12,wisnioski15,wisnioski19,tiley21}. These maps can then be used to measure the rotational velocity $v_{\rm rot}$ and intrinsic velocity dispersion $\sigma_0$ \citep[e.g.,][]{forster-schreiber09,wisnioski15,wisnioski19,johnson18}. In the local Universe there are several comprehensive studies of galaxy kinematics, with surveys such as the Calar Alto Legacy Integral Field Area \citep[CALIFA;][]{sanchez12}, the Sydney--Australian--Astronomical Observatory Multi-Object Integral-field Spectrograph \citep[SAMI; ][]{croom12} and Mapping Nearby Galaxies at Apache Point Observatory \cite[MANGA; ][]{bundy15} providing IFU observations of the gas and stellar motions in thousands of $z$\,$\sim$\,0 galaxies spanning a range of stellar masses.

At $z$\,$\sim$\,2, the rest frame-optical nebular emission lines such as H$\alpha$ and [O{\sc iii}] are redshifted into the near-infrared (NIR) and into the coverage of instruments such as the $K$-band Multi-Object Spectrograph \citep[KMOS;][]{sharples13}.
However, dynamical analyses with KMOS at this epoch are challenging because of the seeing-limited spatial resolution -- KMOS achieves a resolution of $\sim$\,0.6$''$ (FWHM), which corresponds to a physical size of $\sim$\,5\,kpc at $z=$\,2. Nevertheless, with a sufficient signal-to-noise ratio (S/N), and exploiting velocity gradients one can centroid the emission in independent velocity channels and thus sample finer spatial scales than implied by the PSF.
In addition the $H$ band, which covers the redshifted H$\alpha$ emission from galaxies at $z$\,$\sim$\,1.2--1.8, suffers from strong sky contamination \citep{soto16,tiley21} that can be challenging to robustly model and remove.

As a result, the tools used to study kinematics at high redshifts are different to those used at low redshifts. Instead of studying detailed scaling relations, cruder measurements of the ratio of rotational velocity to intrinsic velocity dispersion $v_{\rm rot}/\sigma_0$ have been used \citep{weiner06,newman13,wisnioski15}
in an attempt to characterise the kinematics quantitatively. For example, galaxies with $v_{\rm rot}/\sigma_0>$\,1.5 have been considered rotationally supported \citep[e.g.,][]{stott16,tiley21}, whereas galaxies with $v_{\rm rot}/\sigma_0<$\,1.5 are believed to be dominated by turbulent motions that may indicate an on-going or recent merger \citep[e.g.,][]{alaghband-zadeh12}.

Progress in NIR integral field spectrograph technology has allowed IFU studies of increasing numbers of high-redshift sources in recent years, and as in the local Universe there are now several large surveys of spatially resolved kinematics with KMOS and SINFONI including the Spectroscopic Imaging survey in the near-infrared with SINFONI \citep[SINS/zC-SINF;][]{forster-schreiber09,mancini11}, the KMOS Redshift One Spectroscopic Survey \citep[KROSS;][]{stott16}, the KMOS$^{\rm 3D}$ survey \citep{wisnioski15,wisnioski19}, the KMOS Deep Survey \citep[KDS;][]{turner17} the KMOS Galaxy Evolution Survey \citep[KGES;][]{tiley21} and the KMOS Lensed Emission Lines and VElocity Review \citep[KLEVER;][]{curti20}. These surveys bracket the epoch when the star-formation rate density (SFRD) is at its peak, $z$\,$\sim$\,1--2, and when a significant proportion of the stellar mass we see in the local Universe was assembled. Results from these surveys have revealed that high-redshift star-forming galaxies appear dynamically ``hot'' when compared to local galaxies \citep[e.g.,][]{forster-schreiber09,wisnioski15,wisnioski19,stott16,johnson18}.

Kinematic surveys also enable another probe of galaxy evolution: the Tully-Fisher relation \citep[TFR;][]{tully1977} -- the relationship between galaxy luminosity and rotational velocity -- which can trace the evolution of star-forming galaxies between different epochs. The TFR has been well studied at $z$\,$\sim$\,0 \citep{tully00, lagattuta13}. Surveys at $z$\,$\sim$\,2 find much greater scatter in the relation potentially due to the increased turbulence in the star-forming galaxy population \citep[e.g.,][]{gnerucci11}. This also may be at least partially attributed to lower data quality and to different methods of defining the rotation velocity. These studies have made conflicting claims about the evolution of the TFR. Some find no evolution \citep[e.g.,][]{conselice05,kassin07, miller11,miller12,diteodoro16,pelliccia17,tiley19}, whereas others find evidence for an evolution of the normalisation with redshift \citep[e.g.,][]{cresci09,swinbank12b,ubler17}.

In contrast to these studies of typical star-forming galaxies, spatially resolved kinematic studies of the more active and dust-obscured  DSFGs however, have been much more limited in scope. Among the few published studies are \cite{swinbank06b}, who used the UKIRT Imaging Spectrometer UIST and found four of their sample of six DSFGs to contain multiple components and \cite{alaghband-zadeh12}, who observed nine DSFGs at $z$\,$\sim$\,2.0--2.7 with SINFONI and the Gemini-North/Near-Infrared Integral Field Spectrograph (NIFS), measuring an average H$\alpha$ velocity dispersion of $\sigma$\,$=$\,220\,$\pm$\,80\,km\,s$^{-1}$ indicating high turbulence in these sources. Additionally, they found that six of the nine sources showed multiple kinematically distinct components, and they classified all nine sources as mergers based on kinemetry of the velocity and velocity dispersion maps.
Similarly, \cite{menendez-delmestre13} observed three DSFGs with the OH-Suppressing Infrared Imaging Spectrograph (OSIRIS) on the Keck telescope, finding the systems to contain multiple clumps that they suggested to be in the process of merging, and thus driving high SFRs. More recently, \cite{olivares16} observed eight DSFGs at $z$\,$\sim$\,1.3--2.5 with SINFONI, finding irregular/clumpy velocity and velocity dispersion fields, which they also interpreted as evidence for galaxy-galaxy interactions and/or mergers.

Studying the kinematics of high-redshift DSFGs is one of the main goals of our KMOS Large Programme ``KMOS+ALMA Observations of Submillimetre Sources'' (KAOSS). KAOSS targets $\sim$\,400 DSFGs with KMOS in the $HK$ filter, which covers the H$\alpha$ and/or [O{\sc iii}] emission lines at $z$\,$\sim$\,1--3, and in this paper we will utilise KAOSS to map the H$\alpha$ emission in the brightest, most extended sources, from which we will extract velocity fields and rotation curves. Our goal is to significantly increase the sample of DSFGs with spatially resolved measurements of rotational velocities and intrinsic velocity dispersions, along with $v_{\rm rot}/\sigma_0$, the latter of which is a key diagnostic of the level of rotational support in galaxies \citep[e.g.,][]{wisnioski19}. Our analysis is based on data taken in the first half of the survey \citep[see also][]{birkin_thesis}. The full survey will be presented in Taylor et al.\ (in prep.).

In this paper we present the properties of a subset of 27 KAOSS sources in the COSMOS, UDS and GOODS-S fields, with sufficiently bright and extended H$\alpha$ detections to yield robust 2-D kinematic information. The outline of this paper is as follows: in \S\ref{sec:sample_sel_obs} and \S\ref{sec:data_reduction_analysis} we describe the sample studied and the observations carried out, along with our data reduction and analysis methods, before discussing the measurements made. In \S\ref{sec:results_discussion} we discuss the results and their implications. In \S\ref{sec:conclusions} we summarise our findings. Throughout this paper we adopt the cosmology measured by \cite{planck20} i.e. flat with $\Omega_\mathrm{m}$\,$=$\,0.310 and H$_0$\,$=$\,67.7\,km\,s$^{-1}$\,Mpc$^{-1}$.

%
%
\section{Sample selection and observations}
\label{sec:sample_sel_obs}

%
%
\subsection{Sample}
\label{sec:sample}

This paper uses KMOS data taken for the KAOSS Large Programme (Programme ID: 1103.A-0182). These are 13.5-ks exposure observations of DSFGs in the $HK$ grating ($\sim$\,1.4--2.4\,$\mu$m) with KMOS on the Very Large Telescope (VLT), designed to obtain NIR redshifts and spatially resolved emission-line detections. The KAOSS targets were selected from four ALMA surveys:

{\bf AS2UDS} \citep{stach19}: ALMA 870\,$\mu$m follow up of 712 850\,$\mu$m sources from a deep 0.9 sq.\ degree SCUBA-2 map of the UKIDSS Ultra Deep Survey (UDS) field \citep{geach17}.

{\bf AS2COSMOS} \citep{simpson20}: ALMA 870\,$\mu$m pilot follow up of the 160 brightest 850\,$\mu$m sources from a deep 2 sq. degree SCUBA-2 map of the Cosmic Evolution Survey (COSMOS) field \citep{simpson19}.

{\bf ALESS} and {\bf BASIC} \citep{hodge13,cowie18}: ALMA follow up of 179 LABOCA or SCUBA-2 sources in the GOODS-S/ECDFS field.

{\bf A3COSMOS} \citep{liu19a}: pipeline exploiting the ALMA archive to locate submillimetre-detected galaxies in the COSMOS field.

When selecting targets for the KMOS IFUs we prioritise sources that are brighter in the $K$ band, as they are more likely to yield emission-line detections in our 13.5\,ks exposures (see \S\ref{sec:obs_strategy}), and also more likely to yield resolved kinematics.

For this paper, we select KAOSS sources with H$\alpha$ detections that are bright enough to search for resolved velocity structure from the H$\alpha$ emission line. All sources with line detections from KAOSS are fit on a spaxel-by-spaxel basis, as we will describe in \S\ref{sec:vel_maps}. We consider a source to be kinematically ``resolved'' if the fitting successfully reproduces a velocity map which is extended along the major axis by more than twice the FWHM of the PSF.
This results in a sample of 27 H$\alpha$ sources with $f_{{\rm H}\alpha}>$\,1.7\,$\times$\,10$^{-20}$\,W\,m$^{-2}$ at $z$\,$=$\,1.3--2.6, all of which have a signal-to-noise ratio of S/N\,$>$\,7 for the H$\alpha$ emission line in the integrated spectra.

%
%
\subsection{Physical properties and comparison with other survey samples}
\label{sec:physical_props}

\begin{figure*}
    \centering
    \includegraphics[width=0.48\linewidth]{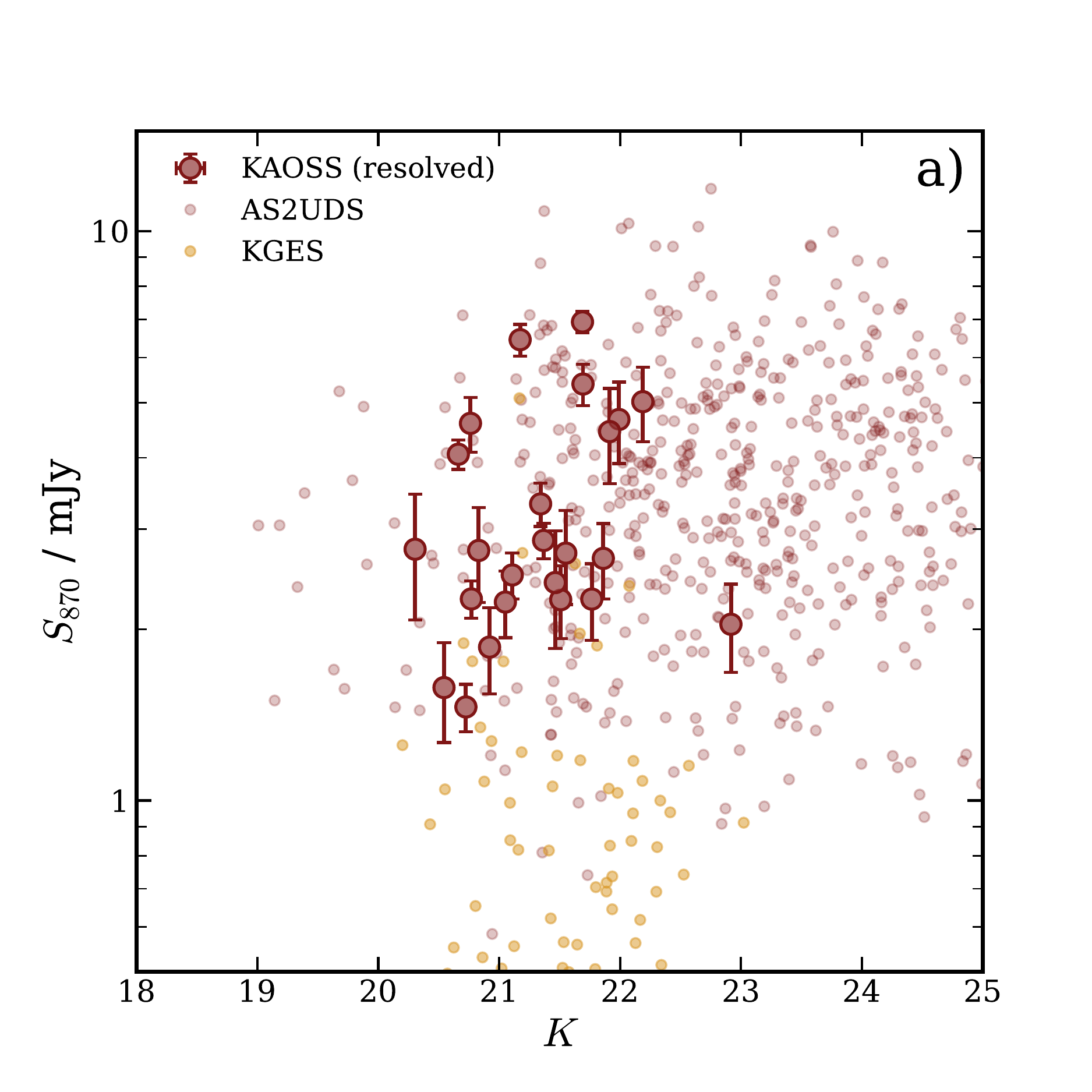}
    \includegraphics[width=0.48\linewidth]{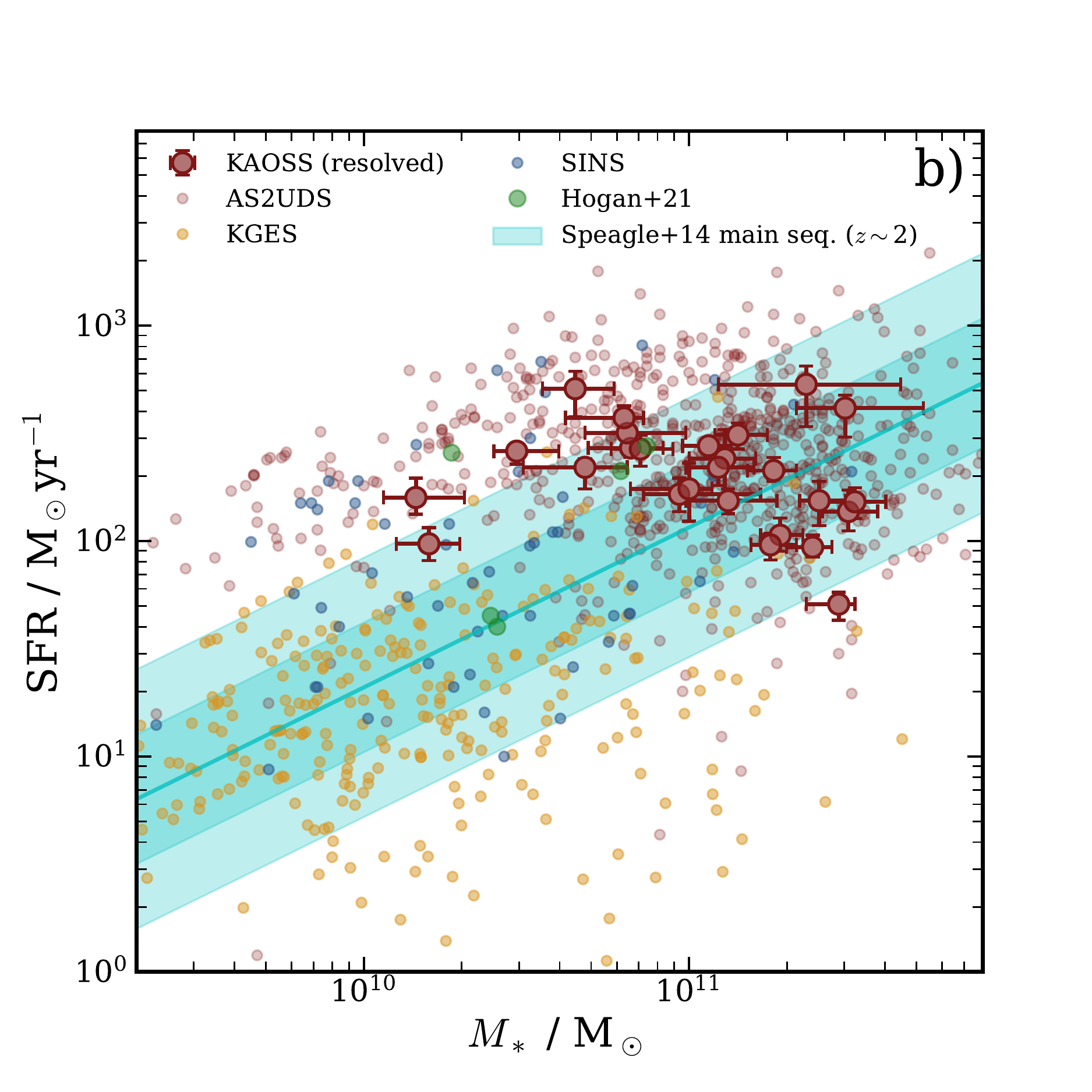}
    \caption{{\bf a)} $K$-band magnitude versus 870-$\mu$m flux density ($S_{870}$) for the KAOSS resolved sample compared with the 707 AS2UDS DSFGs \protect\citep{dudzeviciute20} and $z$\,$\sim$\,1.5 normal star-forming galaxies from KGES \protect\citep{tiley21}. The KAOSS resolved sample is generally representative of the range of 870-$\mu$m flux densities in the DSFG population, but biased towards near-infrared-brighter sources. The KAOSS sample is complemented by KGES in that the latter extends the range in star-formation rate by an order of magnitude. {\bf b)} Star-formation rate versus $M_\ast$ for the same samples, now also including six Herschel-selected $z$\,$\sim$\,2.5 ULIRGs from \protect\cite{hogan21}. KAOSS probes sources that are approximately an order of magnitude more massive than KGES, and overlaps with the parameter range of the Herschel-selected ULIRGs in \protect\cite{hogan21}. Also shown here are $z$\,$\sim$\,1.3--2.6 UV/optically selected star-forming galaxies from the SINS survey \protect\citep{forster-schreiber09}, which has some overlap with KAOSS, but also probes lower-mass and lower-SFR systems. We show the star-forming main sequence according to the prescription of \protect\cite{speagle14}, with the cyan filled regions showing factors of two and four spread in SFR.
    }
    \label{fig:kaoss_selection}
\end{figure*}

Before discussing the resolved kinematics, we place our sample in context with samples from other galaxy surveys. In Fig.~\ref{fig:kaoss_selection} we show the 870-$\mu$m fluxes of the KAOSS sample versus their $K$-band magnitudes. As a comparison sample we show the 707 DSFGs from the ALMA-SCUBA-2 Ultra Deep Survey \citep[AS2UDS;][]{stach19,dudzeviciute20}, which is the largest sample of 870\,$\mu$m-selected DSFGs and therefore provides a good indicator of the properties of the general 870\,$\mu$m-selected population. Our KAOSS sample spans the range of 870-$\mu$m fluxes in AS2UDS, $S_{870}$\,$\sim$\,0.5--14\,mJy, although the resolved subset only samples sources with $K\lesssim$\,23.

Before comparing to other kinematic surveys we first note here that in what follows we mostly relate our results to those from other H$\alpha$ studies. There have been a number of kinematic studies of DSFGs and less active populations carried out using various far-infrared or (sub)millimetre emission lines including CO, [C{\sc i}] and [C{\sc ii}] \citep[e.g.,][]{hodge12,lelli18,lelli21,rizzo20,rizzo21,rizzo23,fraternali21,posses23,roman-oliveira23}, which trace different phases of the gas, however it has been noted by several authors that velocity dispersions measured from CO/[C{\sc i}]/[C{\sc ii}] appear systematically lower than those measured from ionised gas tracers in the optical and near-infrared \citep[e.g.,][]{levy18,su22,lelli23}. This systematic difference would complicate any comparisons, and therefore we generally only consider other H$\alpha$ studies in this work.

We primarily aim to compare with more typical and hence less actively star-forming galaxies, and so we also show in Fig.~\ref{fig:kaoss_selection} $z$\,$\sim$\,1.5 $K$-band-selected star-forming galaxies from the KMOS Galaxy Evolution Survey \cite[KGES;][]{tiley21}. 870-$\mu$m fluxes are not available for these sources, but they do have {\sc magphys}-derived dust mass estimates, which we convert to $S_{870}$ estimates using the $M_{\rm dust}$--$S_{870}$ relation derived by \cite{dudzeviciute20}. These are therefore approximate values, but they highlight the region of the parameter space probed by the KGES sample. In general, KGES galaxies have lower dust masses and hence submillimetre fluxes than KAOSS, but comparable rest-frame optical fluxes.

In Fig.~\ref{fig:kaoss_selection} we also show the distribution of our sample in terms of star-formation rates (SFRs) versus stellar masses ($M_\ast$) taken from pre-existing {\sc magphys} \citep{dacunha15, battisti19} spectral energy distribution (SED) fits. For details on the SED fitting for these sources we direct the reader to \cite{dacunha15}, \cite{dudzeviciute20} and Ikarashi et al.\ (in prep.). These fits are performed using the photometric redshift extension of the {\sc magphys} code, and therefore the redshifts are probabilistic in the modelling at this point. In \S\ref{sec:magphys} we will repeat the fits for sources with spectroscopic redshifts from KAOSS.
We also show the star-forming main sequence at $z$\,$=$\,2 according to the prescription of \cite{speagle14}. The 27 resolved KAOSS sources have median stellar masses and star-formation rates of $M_\ast$\,$=$\,(1.3\,$\pm$\,0.2)\,$\times$\,10$^{11}$\,M$_\odot$ and SFR\,$=$\,220\,$\pm$\,30\,M$_\odot$yr$^{-1}$. To compare this with other DSFGs, we select the 283 AS2UDS sources lying in the range $z=$\,1.3--2.6, encompassing all 27 resolved KAOSS sources. These have median stellar masses and SFRs of $M_\ast$\,$=$\,(1.44\,$\pm$\,0.01)\,$\times$\,10$^{11}$\,M$_\odot$ and SFR\,$=$\,173\,$\pm$\,6\,M$_\odot$yr$^{-1}$, hence in this analysis we are probing DSFGs that are generally representative of the stellar masses in the 870\,$\mu$m-selected population, but slightly more active in terms of star-formation rate.

Fig.~\ref{fig:kaoss_selection} demonstrates that compared to KAOSS the KGES sample probes much less massive and less actively star-forming sources, with median stellar masses and SFRs of $M_\ast$\,$=$\,(1.3\,$\pm$\,0.1)\,$\times$\,10$^{10}$\,M$_\odot$ and SFR\,$=$\,16\,$\pm$\,1\,M$_\odot$yr$^{-1}$, respectively, approximately an order of magnitude lower than the KAOSS resolved sample in both cases. Therefore, by supplementing our results with those from KGES we will be able to study the variation of kinematic properties across a wider range in stellar mass and star-formation rate.

As a further comparison sample of similar galaxies, in Fig.~\ref{fig:kaoss_selection} we include data from KMOS observations of six $z$\,$\sim$\,2.5 Herschel-selected ULIRGs with kinematical information estimated by \cite{hogan21}. While these sources are selected based on shorter far-infrared wavelengths than our DSFGs, they are gas-rich star-forming galaxies at comparable redshifts to the most distant sources in our resolved sample, which spans the range $z$\,$\sim$\,1.3--2.6. The six sources from \cite{hogan21} have median stellar masses and SFRs of $M_\ast$\,$=$\,(2.5\,$\pm$\,1.5)\,$\times$\,10$^{11}$\,M$_\odot$ and SFR\,$=$\,130\,$\pm$\,90\,M$_\odot$\,yr$^{-1}$, consistent with the KAOSS resolved sample. Where possible we compare our results with both the KGES and ULIRG samples (along with other samples of less active galaxies) throughout our analysis.

%
%
\subsection{Observing strategy}
\label{sec:obs_strategy}

KMOS \citep{sharples13} is a near-infrared multi-object spectrograph mounted at the Nasmyth focus of Unit Telescope 1 on the VLT. It is comprised of 24 IFUs that patrol a field of 7.2$'$ diameter area of the sky, with each IFU covering a 2.8$''$\,$\times$\,2.8$''$ field of view sampled by 14\,$\times$\,14 spaxels (0.2$''$ per pixel). In our survey fields KMOS pointings typically contain $\sim$\,10 DSFGs, and as KMOS has 24 IFUs available, and given the need for sky offsets, we choose to pair IFUs on our targets and a matched blank sky region where possible. By creating sky positions offset relative to the corresponding target position by a similar fixed vector, we can ensure that the target is observed by either the primary or secondary (sky) IFU at all times. When observing, the instrument nods back and forth between the target and a sky position in order to assist sky subtraction. This maximises the on-target time.

The final reduced frames include the target in both the primary or secondary IFUs which are then combined to increase the signal-to-noise ratio (S/N). Pairing of these IFUs is not {\it always} possible given the positioning of targets within the field of view, but we prioritise this approach for sources with $K$\,$<$\,22.5 that are most likely to yield spatially resolved kinematics. We also reserve one IFU to be placed on a bright ($H\sim$\,12--15) star, allowing us to monitor the telescope pointing and the point spread function (PSF). The median PSF FWHM of our observations is 0.59\,$\pm$\,0.06$''$ where the quoted uncertainty is the standard deviation.

Each observing block (OB) yields 2.7\,ks of on-source integration time in around an hour of telescope time. To obtain our desired sensitivity of $\sim$\,1\,$\times$\,10$^{-20}$\,W\,m$^{-2}$ we observe each OB five times, resulting in a total exposure time of 13.5\,ks for each pointing. Observations are carried out in the combined $HK$ grating, which covers the wavelength range $\lambda\sim$\,1.4--2.4\,$\mu$m at a spectral resolution of $\lambda$/$\Delta\lambda$\,$\sim$\,2000 (corresponding to an instrumental $\sigma$ of $\sim$\,63\,km\,s$^{-1}$). This wavelength range covers the H$\alpha$ emission line for sources at $z$\,$\sim$\,1.1--2.7 where the majority of our targets are expected to reside, given their photometric redshifts.

%
%
\section{Data reduction and analysis}
\label{sec:data_reduction_analysis}

%
%
\subsection{KMOS data reduction}
\label{sec:data_reduction}

In this section we provide a brief description of the processes taken to produce fully reduced data cubes from the raw KMOS data, following the approach used by \cite{tiley21}. More details are provided in \cite{birkin_thesis}.

Calibration of the raw data products proceeds via the European Southern Observatory (ESO) Recipe Execution Tool \citep[{\sc esorex};][]{esorex}, a library of functions that take as an input the raw data and produce reduced 3D cubes. While the standard {\sc esorex} pipeline carries out a basic A$-$B sky subtraction, this is often poor in the $HK$ band of KMOS, therefore we employ a more sophisticated technique based on the Zurich Atmospheric Purge \citep[ZAP;][]{soto16} method initially developed for the MUSE instrument, with optimisations made for KMOS observations. The ZAP method is based on principal component analysis (PCA), using filtering and data segmentation to reduce sky emission residuals while preserving flux from the astronomical target. This KMOS-adapted method is encapsulated in the {\sc pyspark} code (Mendel et al.\ in prep.).

As previously stated, in every OB we assign at least one IFU to a bright star, one of the primary purposes of which is to centre the data cubes between OBs. Therefore, for each set of AB pairs we obtain a reduced cube of a bright star, and we measure the centroid of the emission in this cube by collapsing it and fitting a 2-D Gaussian profile to the spatial emission. We then shift all observations of that field to a common centre using the measured centroid from the star. We also check for any significant offset in the final cubes between individual observations. This allows us to be confident that we are not losing S/N in our combined cubes, as a result of misalignment. Small perturbations can affect the alignment of the telescope, and while in theory these should be corrected for in the acquisition and data reduction, we check each observing block ($\sim$\,1\,hr exposure) for offsets by comparing the measured position of any sources bright enough in their continuum or line emission to be detected. Once the reduced cubes have been produced and aligned, we stack each source individually by taking the mean of each frame and applying a 3$\sigma$ clip.

%
%
\subsection{Spectral extraction and line identification}
\label{sec:spectra_line_IDs}

Having reduced the KMOS data we ``unwrap'' the cubes into 2-D spectra, noting any potential line emission and cross-referencing with pre-existing photometric and/or spectroscopic redshifts to assist in identifying the emission lines. For the 27 sources in our preliminary resolved sample, photometric and spectroscopic redshifts were available for 22 and 11 sources, respectively. For sources where we believe a line to be present we collapse the cube around the approximate wavelength of the observed emission line and visually inspect the resultant line map, then extract a 1-D spectrum at the position of the emission in an aperture of radius 0.6$''$.

%
%
\subsection{H$\alpha$ velocity and velocity dispersion maps}
\label{sec:vel_maps}

To determine the kinematics of our sources we model the H$\alpha$ emission line in each spaxel. By fitting the emission line we can derive resolved maps of the velocity and velocity dispersion from which we will extract rotation curves and measure rotational velocities.

\begin{figure*}
    \centering
    \includegraphics[width=1\linewidth]{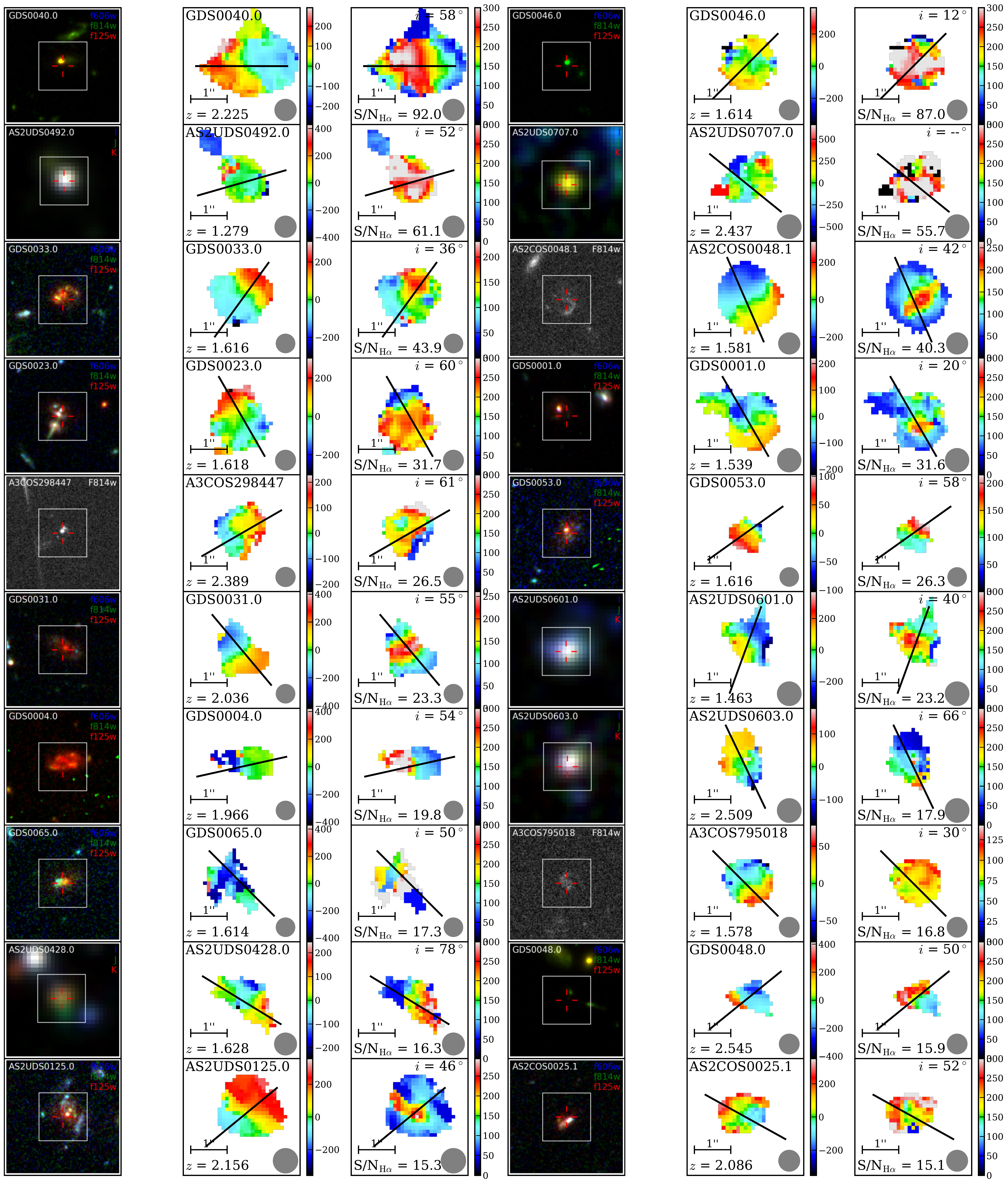}
    \caption{HST or ground-based colour images (left), velocity fields (middle) and velocity dispersion maps (right) for our sample of 27 resolved KAOSS DSFGs. The images are 7.5$''$\,$\times$\,7.5$''$ in size, and the white box indicates the region shown in the velocity and velocity dispersion maps (3.2$''$\,$\times$\,3.2$''$ in size). Alongside the maps we also show a 1$''$ scale marker in the bottom-left corner, and the corresponding PSF FWHM for the observations of each target in the bottom-right corner. Sources are ordered by the S/N of the H$\alpha$ emission line (shown in the bottom of the right-hand panels) and we indicate the inclination angle derived for the source from {\sc galfit} modelling (\S\ref{sec:galfit}). The black lines on the centre and right panels indicate the kinematic position angle along which rotation curves are measured. For the left panels we indicate the three filters that make up the RGB colour image, or the single filter in cases where the image is greyscale.}
    \label{fig:kaoss_vel_fields_1}
\end{figure*}

\renewcommand{\thefigure}{\arabic{figure} (Cont.)}
\begin{figure*}
    \ContinuedFloat
    \captionsetup{list=off,format=cont}
    \centering
    \includegraphics[width=1\linewidth]{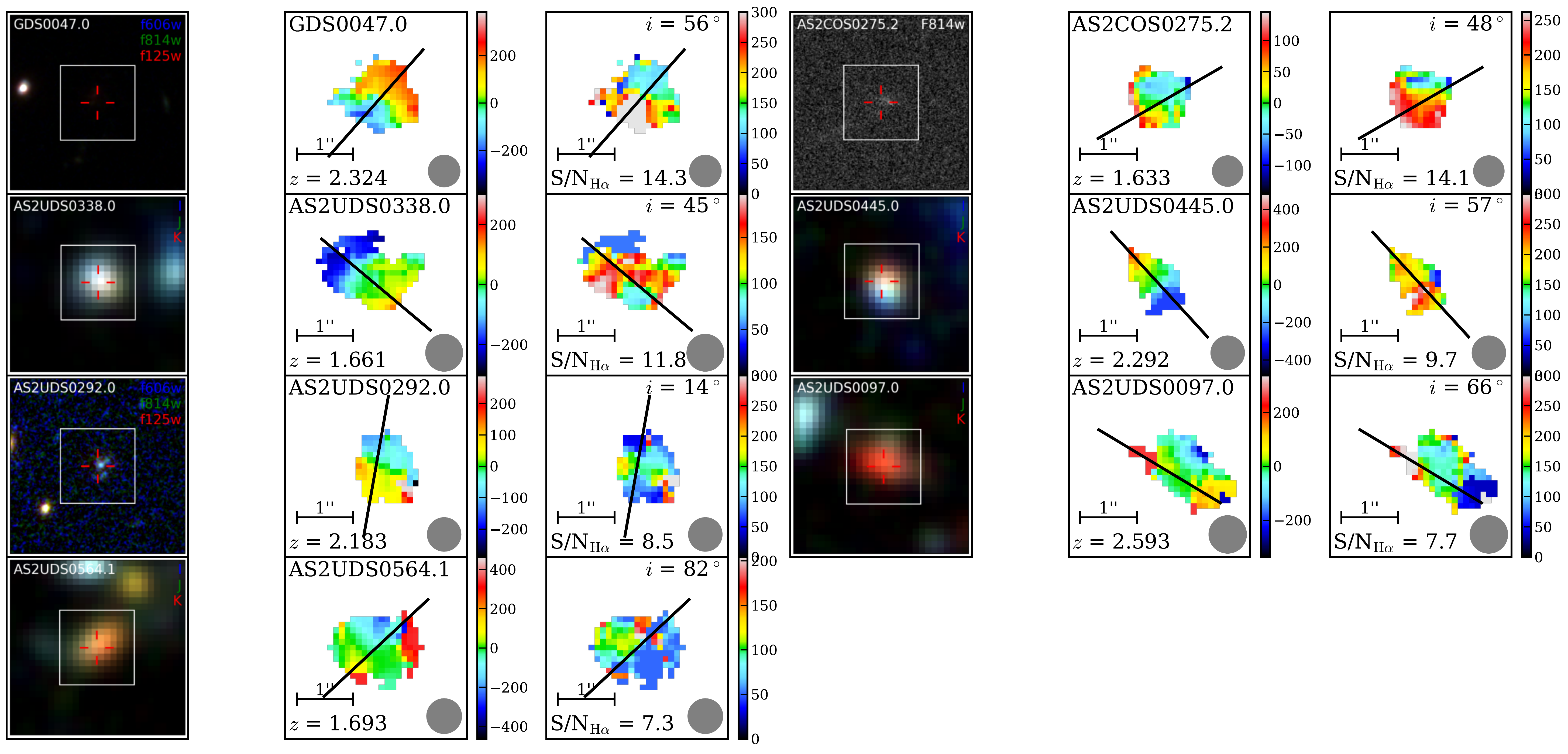}
    \caption{}
    \label{fig:kaoss_vel_fields_2}
\end{figure*}
\renewcommand{\thefigure}{\arabic{figure}}

We fit a three-component Gaussian profile, plus a constant continuum component, to the H$\alpha$ line and [N{\sc ii}]$\lambda\lambda$6548,6583 doublet, coupling their wavelengths and linewidths, with the [N{\sc ii}]$\lambda6583$/H$\alpha$ flux ratio as a free parameter, and fixing the [N{\sc ii}]$\lambda6583$/[N{\sc ii}]$\lambda6548$ flux ratio to a value of 2.8 \citep{osterbrock06}. We perform the fitting over the region of the spectra within $\pm$\,0.02\,$\mu$m of the H$\alpha$ emission line. Observed linewidths are deconvolved with the instrumental resolution, as determined by fitting several sky lines over the $HK$ band, to calculate the intrinsic linewidths.

We perform the fitting on a pixel-by-pixel basis, first resampling the velocity fields from a spatial pixel scale of 0.2$''$ to 0.1$''$, enabling us to sample finer spatial scales. For each pixel we attempt to fit the emission lines, and if the fit does not achieve a threshold of S/N\,$=$\,5 we bin with neighbouring pixels, increasing the bin size and repeating up to a maximum bin size of 5 pixels (0.5$''$) or until the S/N threshold is achieved. For the systemic redshifts we use the values derived from fits to the integrated spectra \citep{birkin_thesis}. We note here that given the complex kinematic structure and dust obscuration seen in some of our sources, this is not always a good indicator of the true kinematic centre (see \S\ref{sec:rot_curve_modelling}). 

Velocity and velocity dispersion maps for all the galaxies in the resolved sample are shown in Fig.~\ref{fig:kaoss_vel_fields_1} alongside rest-frame optical colour images of the sources. For the colour images we include high-resolution HST imaging where possible, otherwise we use ground-based imaging. Sources are ordered by the integrated S/N of the H$\alpha$ emission line which generally correlates with the quality of the kinematic information from the fitting. Several of the sources in Fig.~\ref{fig:kaoss_vel_fields_1} display smooth H$\alpha$ velocity gradients, such as AS2COS0048.1, GDS0033.0 and GDS0031.0, which indicate ordered rotation in these galaxies. Others, such as GDS0001.0, GDS0046.0 and AS2UDS0428.0, display more complex velocity structures and morphologies.

Two other sources are worthy of comment. AS2UDS0492.0 is unique in our sample, since we detect velocity structure from a separate component to the north west within the 2.8$''$ field of the KMOS IFU -- this component is detected in the ground-based near-infrared imaging (see Fig.~\ref{fig:kaoss_vel_fields_1}) and corresponds to a companion galaxy. This particular source also displays a broad component in the H$\alpha$ emission, AGN-like IRAC colours and an X-ray component (see \S\ref{sec:agn}). The fit to the H$\alpha$ emission in AS2UDS0707.0 is poor despite the high S/N of the integrated emission (S/N$_{{\rm H}\alpha}\sim$\,56). Like AS2UDS0492.0, this source displays broad H$\alpha$ emission, but it also has a very low [N{\sc ii}]/H$\alpha$ flux ratio unlike what we would expect from an AGN-dominated source.

To illustrate the variety of velocity structures in our sample we also show in Fig.~\ref{fig:kaoss_pv_diagrams_major} position-velocity (PV) diagrams for all 27 KAOSS galaxies. PV diagrams are extracted from a 0.5$''$ wide slit along the kinematic major axis, and are smoothed using a Gaussian window of FWHM corresponding to the seeing in the $x$-direction (spatial) and the velocity resolution in the $y$-direction (velocity). Sources for which we were able to extract kinematic information (the {\it disc-like} sample; see \S\ref{sec:rot_curve_modelling}) are shown in colour, and other sources are greyed out as they are not included in the majority of our analysis. PV diagrams are centred using the results of the Freeman disk fitting described in \S\ref{sec:rot_curve_modelling}. As a check of the data quality, we also generated minor-axis PV diagrams and visually inspected these to confirm that the majority of the H$\alpha$ emission lies close to the systemic redshift and spatial centre.

The velocity dispersion maps in Fig.~\ref{fig:kaoss_vel_fields_1} generally appear structured, even for the sources with the most significant detections, for example AS2UDS0707.0, AS2COS0025.1 and AS2UDS0338.0. Given the compact sizes of our sources relative to the KMOS PSF (FWHM\,$\sim$\,0.6$''$) our observations are susceptible to the effects of beam smearing, leading to increased observed velocity dispersions in the centres of galaxies and flattening of their observed rotation curves \citep[e.g.,][]{johnson18}. However, it can be seen in sources such as AS2COS0048.1 and AS2UDS0125.0 that we resolve the H$\alpha$ emission on scales large enough to reach the point where the velocity dispersion profile flattens out. We provide a discussion of beam smearing in our sample and the methods with which we account for it in \S\ref{sec:beam_smearing}.

%
%
\subsection{JWST/NIRCam imaging}
\label{sec:nircam}

\begin{figure*}
    \centering
    \includegraphics[width=\linewidth]{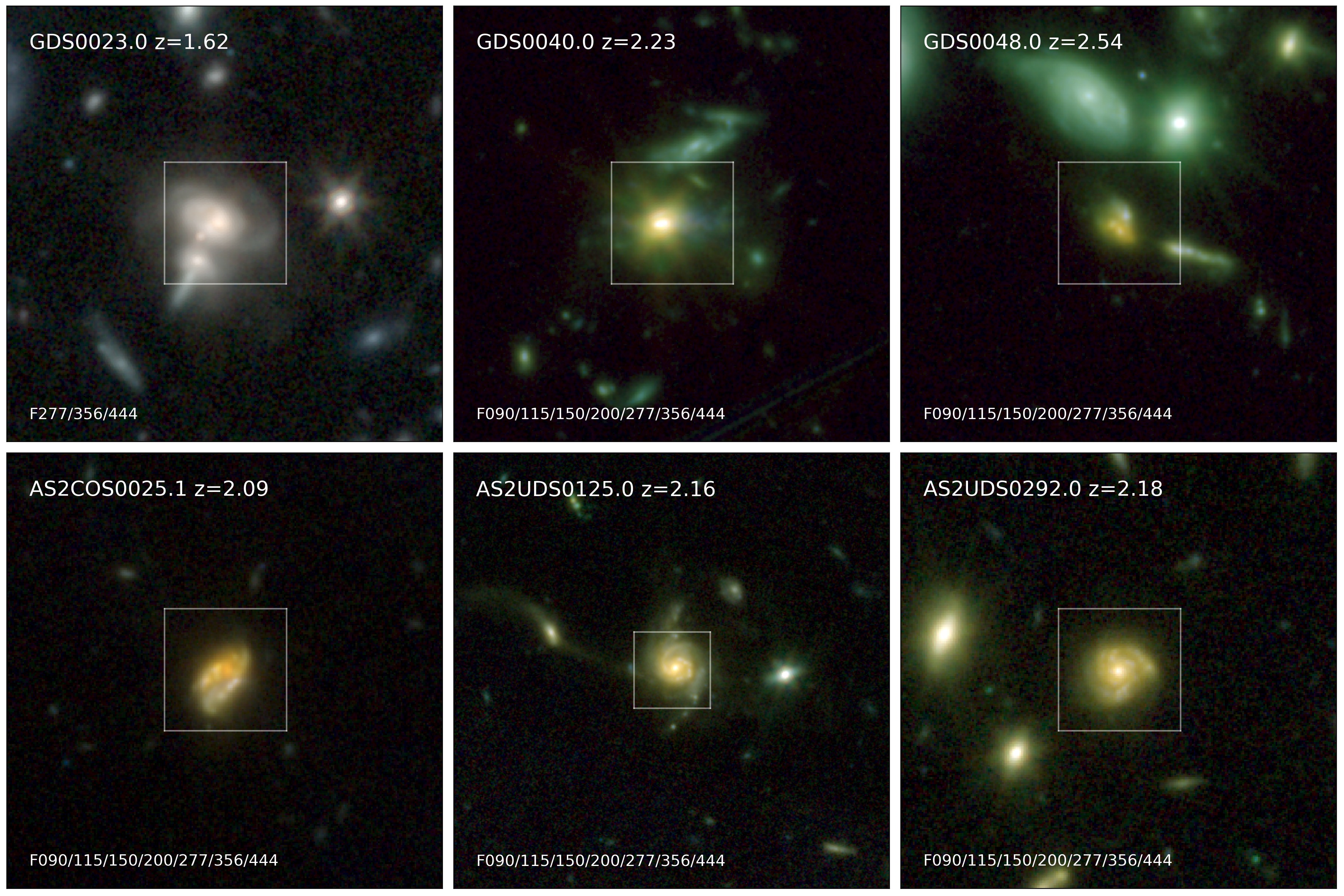}
    \caption{10$''$\,$\times$\,10$''$ (except for AS2UDS0125.0, which is 16$''$\,$\times$\,16$''$) colour images of six KAOSS sources using various combinations of JWST/NIRCam filters F090W, F115W, F150W, F200W, F277W, F356W and F444W. The white boxes indicate the KMOS field of view (2.8$''$\,$\times$\,2.8$''$). The sources display a range of morphologies: AS2UDS0292.0 is disc-like and apparently fairly isolated; GDS0040.0 and AS2COS0025.1 are more morphologically complex. GDS0023.0, GDS0048.0 and AS2UDS0125.0 all appear to be interacting with companion galaxies on projected scales of $\sim$\,2--5$''$ ($\sim$\,16--40\,kpc).}
    \label{fig:nircam}
\end{figure*}

Six of the sources in the resolved sample have been recently imaged with JWST/NIRCam, which represents a significant improvement on the wavelength coverage and sensitivity to the underlying stellar population for obscured galaxies compared to the existing HST imaging. Therefore, in Fig.~\ref{fig:nircam} we show colour images of these six sources \citep[JADES proposal ID: 1180, PI: Eisenstein;][PRIMER proposal ID: 1837, PI: Dunlop]{rieke23}. In all sources we construct blue, green and red channels from the (F090W$+$F115W$+$F150W$+$F200W), (F277W$+$F356W) and F444W filters respectively. The only exception is GDS0023.0, which only has coverage in F227W, F356W and F444W which make up the blue, green and red channels, respectively.

In Fig.~\ref{fig:nircam} we see a range of morphologies, with GDS0023.0, AS2COS0025.1, AS2UDS0125.0 and AS20292.0 showing potential spiral arms. However, several (GDS0023.0, GDS0048.0 and AS2UDS0125.0) appear to be interacting with potential companion galaxies on scales of $\sim$\,2--5$''$ ($\sim$\,16--40\,kpc), evidenced by the tidal features in the images. We confirmed in all cases that the ``interacting'' sources are indeed at the same redshift as the DSFG. The source to the west of GDS0048.0 is confirmed by MUSE observations \citep{inami17} to also be at $z$\,$=$\,2.54, and the source to the east of AS2UDS0125.0 that appears to be tidally interacting has a photometric redshift that is consistent with the H$\alpha$ redshift of the DSFG \citep{dudzeviciute20}. The companion to GDS0023.0 is detected in H$\alpha$ at the same redshift in a $\sim$\,100\,hr integration with KMOS by \cite{puglisi23}, identified as GS4-16960.

In addition, as we show in \S\ref{sec:agn} GDS0040.0 and GDS0048.0 host active galactic nuclei that are detectable in their rest-frame optical spectra. This is likely the cause of the unresolved F444W emission seen in GDS0040.0.

%
%
\subsection{Spectral energy distribution fitting}
\label{sec:magphys}

To derive physical properties for our sources we fit their SEDs employing the high-redshift version of the {\sc magphys} code \citep{dacunha15}, fixing the redshift to the KMOS spectroscopic value. We direct the reader to \cite{birkin21} for a description of our methods for SED fitting with {\sc magphys}, but we note here that, of the 27 sources in the resolved sample, 26 (96 per cent) have Spitzer/MIPS 24-$\mu$m detections and 25 (93 per cent) have at least one Herschel/SPIRE detection.
The photometry used for each of the three fields is as follows:
\begin{itemize}
    \item {\bf COSMOS:} CFHT Megacam $U$, Subaru SuprimeCam $BVRIz$, Subaru HSC $Y$, VISTA $HK_s$, Spitzer IRAC1-4 \citep{laigle16}, Spitzer MIPS 24$\mu$m, Herschel PACS 100$\mu$m, 160$\mu$m, Herschel\\ SPIRE 250$\mu$m, 350$\mu$m, 500$\mu$m \citep{jin18}, ALMA 870$\mu$m\\ \citep{simpson20} and VLA 3\,GHz \citep{smolcic17}
    \item {\bf UDS:} CFHT Megacam $U$, Subaru SuprimeCam $BVRIz$, VISTA $Y$, UKIRT WFCAM $JHK$ \citep{lawrence07}, Spitzer IRAC1-4, Spitzer MIPS 24$\mu$m \citep{kim11}, Herschel PACS 100$\mu$m,\\ 160$\mu$m, Herschel SPIRE 250$\mu$m, 350$\mu$m, 500$\mu$m \citep{oliver12},\\ ALMA 870$\mu$m \citep{stach19} and VLA 1.4\,GHz \citep{simpson13}
    \item {\bf GOODS-S:} VIMOS $U$, HST ACS F435W, F606W, F775W, F814W, F850LP, HST WFC3 F098M, F105W, F125W, F160W, VLT/HAWK-I $K_s$, Spitzer IRAC1-4 \citep{guo13}, Spitzer MIPS 24$\mu$m \citep{giavalisco04}, Herschel PACS 70$\mu$m, 100$\mu$m, 160$\mu$m, Herschel SPIRE 250$\mu$m, 350$\mu$m, 500$\mu$m \citep{elbaz11}, ALMA 870$\mu$m \citep{cowie18,liu19a} and VLA 1.4\,GHz \citep{miller13}.
\end{itemize}

The observed fluxes or limits and the corresponding best-fit {\sc magphys} SEDs for the 27 KAOSS DSFGs are shown as a figure in the online supplementary material. The {\sc magphys} model SEDs provide good fits to the observed photometry in all cases. We note here, however, that the high-redshift version of {\sc magphys} does not include contributions to the continuum emission from an AGN. The effects of this on DSFG stellar mass estimates from {\sc magphys} have been tested in \cite{birkin21}, where it was found that $M_\ast$ may be modestly overestimated by the fitting code, although we do not expect this to affect our conclusions.

The best-fit SED parameters and their uncertainties are shown in the online supplementary table, but here we report that the 27 resolved KAOSS DSFGs have median values of $M_\ast=(1.3\pm0.2)\times10^{11}$\,M$_\odot$, $M_{\rm dust}=(4.9\pm0.9)\times10^{8}$\,M$_\odot$, SFR\,$=$\,210\,$\pm$\,30\,M$_\odot$\,yr$^{-1}$, $L_{\rm IR}=(2.9\pm0.4)\times10^{12}$\,L$_\odot$ and $A_V=2.01\pm0.19$.

%
%
\subsection{{\sc galfit} modelling}
\label{sec:galfit}

When deriving intrinsic rotational velocities the inclination and size of the galaxy are both important measurements. To do this for the KAOSS galaxies we exploit existing HST/F160W imaging for 13 of the 27 DSFGs, and ground-based $K$-band imaging from: VISTA for COSMOS \citep[seeing $\sim$\,0.8$''$]{mccracken12}, UKIRT WFCAM for UDS sources (Almaini et al.\ in prep., seeing $\sim$\,0.6$''$), and VLT HAWK-I for GOODS-S sources \citep[seeing $\sim$\,0.4$''$]{fontana14}; for the remaining 14. We fit the 2-D continuum with S\'{e}rsic profiles using the {\sc galfit} code \citep{peng10}, constraining the S\'{e}rsic index $n$ to be between 0.5 and 4, and allowing the effective radius ($R_{\rm e}$), axis ratio ($b$/$a$) and position angle (PA) to vary. We visually inspect all fits and flag sources where we are unable to find a model that reproduces the source structure, or where the best-fit parameters are unphysical (for example, effective radii of $\ll$\,1 pixel), although this is only necessary for two sources (GDS0046.0 and AS2UDS0707.0). The median S\'{e}rsic index of the entire sample is $n$\,$=$\,1.00\,$\pm$\,0.16 from the fits with $n$ as a free parameter, i.e. consistent with an exponential profile, and we therefore repeat the fitting fixing $n$\,$=$\,1, following \cite{gullberg19}. All parameters are derived after deconvolution with the PSF.

As a test of the suitability of lower-resolution ground-based $K$-band imaging we compare measurements of $R_{\rm e}$ and $b$/$a$ from ground-based $K$-band and HST/F160W imaging for the 12 sources with coverage in both bands, fitting a fixed $n$\,$=$\,1 profile in both cases. The two are consistent within their uncertainties for these sources, and we suggest that size and $b$/$a$ measurements from ground-based $K$-band imaging are acceptable in the absence of HST/F160W imaging. More detail on the fitting procedure can be found in \cite{birkin_thesis}.

In order to estimate uncertainties on the {\sc galfit} parameters we simulate S\'{e}rsic profiles with known properties at different signal-to-noise ratios. We do this for two cases, one with PSFs comparable to the $K$-band imaging and one comparable to the HST/F160W imaging. Finally, we calculate the dispersion in the measurements at different S/N. As a result of these simulations we elect to adopt a constant 10 per cent uncertainty for all measurements of $R_{\rm e}$ and $b/a$, which we find to be generally conservative for the typical S/N levels of the optical/NIR imaging \citep[see][for more details]{birkin_thesis}.

To estimate inclination angles we use the best-fitting {\sc galfit} parameters as follows:
\begin{equation}
    \cos(i) = \sqrt{\dfrac{(b/a)^2-(b/a)_0^2}{1-(b/a)_0^2}},
\end{equation}
where $(b/a)_0$ accounts for the fact that the discs have a finite thickness -- we adopt $(b/a)_0$\,$=$\,0.2, following \cite{gillman19} and other similar surveys such as KROSS \citep{stott16} and KGES \citep{gillman20}, for consistency. Our sample has a median axial ratio of $b/a$\,$=$\,0.64\,$\pm$\,0.03 and a median derived inclination of $i$\,$=$\,52\,$\pm$\,3$^\circ$, broadly consistent with the prediction of 57$^\circ$ for randomly oriented thin discs \citep{law09}.

%
%
\subsection{Determining rotation axes}
\label{sec:rotation_axes}

To quantify the kinematic structure of our sample, we need to parameterise the dynamics through measurements of the rotational velocity, $v_{2.2R_{\rm d}}$, and velocity dispersion, $\sigma_{\rm obs}$. These quantities can be estimated from the rotation curves and velocity dispersion profiles as extracted from the kinematic maps derived in \S\ref{sec:vel_maps}. First we determine the axes across which our sources have the largest velocity gradient. One way to assess this axis is to use the morphological major axis derived in \S\ref{sec:galfit}, PA$_{\rm morph}$. Alternatively we can use the velocity field itself to estimate a kinematic axis, PA$_{\rm kin}$.

To determine PA$_{\rm kin}$ it is necessary to ensure that the velocity fields are appropriately centred, for which we employ the following method. First, we attempt to measure a centroid from the continuum image of the source constructed from the collapsed KMOS cube. If the continuum is not detected, we next measure a centroid from the H$\alpha$ image. In the event that both of these methods are unsuccessful (i.e. if the H$\alpha$ S/N is low) we visually inspect the velocity field to determine its centre. In total we measure centroids for 21, 3 and 3 sources for the three methods, respectively. We then shift the velocity field to align it with the chosen centroid, corresponding to a median shift of 0.40$''$\,$\pm$\,0.06$''$ ($\sim$\,3.5\,kpc).

\begin{figure}
    \centering
    \includegraphics[width=\linewidth]{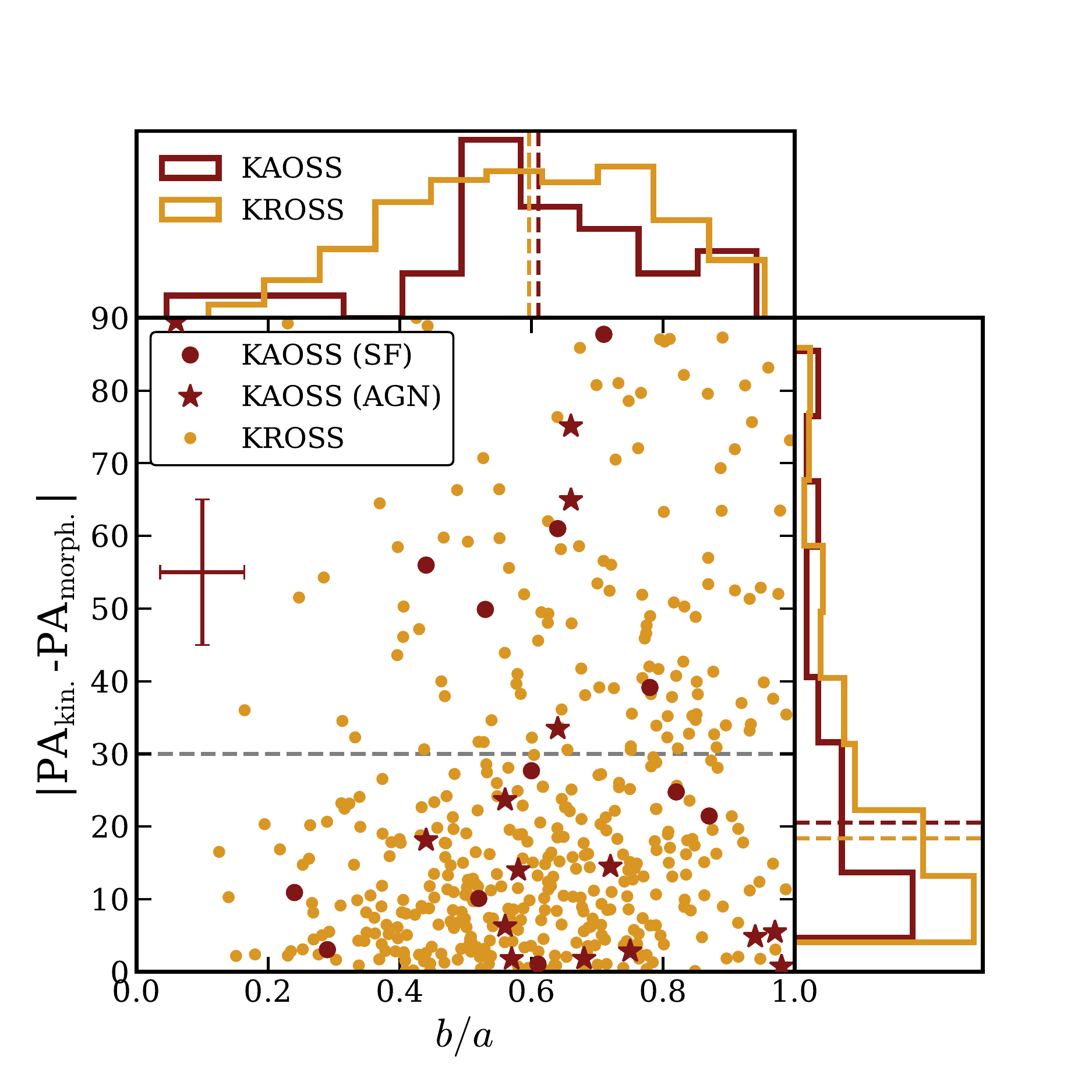}
    \caption{Misalignment between the position angle derived from H$\alpha$ kinematics (PA$_{\rm kin}$) and the position angle derived from optical/NIR imaging (PA$_{\rm morph}$), versus axis ratio ($b/a$). PA$_{\rm morph}$ and $b/a$ are both measured from {\sc galfit} modelling of high-resolution optical/NIR imaging (see \S\ref{sec:galfit}). The horizontal dashed line indicates a misalignment of 30$^\circ$, less than which we take to be reasonable agreement between the two position angles, given the typical uncertainties. In the top-left corner we show a representative error bar for KAOSS galaxies, and as a comparison sample we plot results from the KROSS survey of typical star-forming galaxies at $z$\,$\sim$\,1 \protect\citep{harrison17}. A Kolmogorov-Smirnov test shows that the two distributions are consistent with having been drawn from the same parent sample at the 95 per cent confidence level, indicating that the KAOSS resolved sample shows similar levels of kinematic misalignment to ``main-sequence'' galaxies at $z$\,$\sim$\,1.}
    \label{fig:theta_pa}
\end{figure}

Having centred our velocity fields we then determine their rotational axes (or kinematic position axes; PA$_{\rm kin}$) using two methods. First, we place a pseudo-slit across the velocity field and measure the peak-to-peak difference in velocity, $\Delta v$, then rotate the pseudo-slit through to determine $\Delta v$ as a function of $\theta$, from which we find the angles that both maximise and minimise $\Delta v$: $\theta_{\Delta v_{\rm max}}$ and $\theta_{\Delta v_{\rm min}}$. Finally we use
\begin{equation}
    {\rm PA}_{\rm kin} = \dfrac{\theta_{\Delta v_{\rm max}}+(\theta_{\Delta v_{\rm min}}+90^\circ)}{2}
\end{equation}
to derive the PA$_{\rm kin}$. To estimate uncertainties on the PA$_{\rm kin}$ we employ a Monte Carlo technique, randomly resampling the velocity fields 100 times using the measurement uncertainties, and measuring the spread in the distribution of the resultant 100 values.

When the S/N of the emission is low, this method is noisy and the resulting PA$_{\rm kin}$ may not appear to correlate well with the velocity field. Hence, in all cases we also identify a maximum velocity gradient PA$_{\rm kin}$ by eye, and where the PA$_{\rm kin}$ chosen by the algorithm described above does not appear to be a good fit to the velocity field (which we assess visually) we simply use the ``by eye'' PA$_{\rm kin}$. For context, we use the visual estimate of PA$_{\rm kin}$ for 12 (typically lower S/N) out of the 27 sources in the sample. For these sources we adopt an uncertainty of $\pm$\,5$^\circ$, which is comparable to the median uncertainty estimated from the 15 sources with Monte Carlo-derived uncertainties, as described above. Our best-estimated values of the PA are tabulated in the online supplementary table, and in Fig.~\ref{fig:kaoss_vel_fields_1} we overlay PA$_{\rm kin}$ on both the velocity and velocity dispersion maps.

For galaxies that are highly inclined, the kinematic and morphological position angles should be consistent (i.e. PA$_{\rm kin}$\,$=$\,PA$_{\rm morph}$), however in systems that are closer to face-on or kinematically and morphologically complex this is not necessarily true. Comparing the two position angles as a function of ellipticity therefore provides another metric for identifying disturbed systems \citep[e.g.,][]{wisnioski15,harrison17}. In Fig.~\ref{fig:theta_pa} we show the misalignment between the kinematic position angle PA$_{\rm kin}$ and the morphological position angle measured from the {\sc galfit} modelling of the optical/NIR imaging, PA$_{\rm morph}$ (see \S\ref{sec:galfit}), as a function of the major-to-minor axis ratio $b/a$ (also derived from {\sc galfit}, see \S\ref{sec:galfit}). We indicate a misalignment limit of 30$^\circ$ following \cite{wisnioski15}, finding that 
nine galaxies lie above this threshold and 18 below it. Therefore, 33\,$\pm$\,11 per cent of the resolved KAOSS DSFGs display kinematic and morphological axes that are misaligned by more than 30$^\circ$.

As a comparison sample we also show the distribution of galaxies from the KROSS sample of typical star-forming galaxies at $z$\,$\sim$\,1 \citep{harrison17}, with histograms of the distributions shown on both axes. We perform a two-sample Kolmogorov-Smirmov (K--S) test between the distributions of both the PA offsets and the axial ratios from KAOSS and KROSS, finding them both to be consistent with being drawn from the same parent population at the 95 per cent confidence level. The KROSS sample is comprised of main-sequence star-forming galaxies at $z$\,$\sim$\,1, with typical star-formation rates of $\sim$\,7\,M$_\odot$\,yr$^{-1}$, and this suggests that our sample is no more kinematically complex than much less-active SFGs, in terms of the axial misalignment. Later in our analysis we will further test this result by comparing the star-formation rates and velocity dispersions of different samples.

%
%
\subsection{AGN classification}
\label{sec:agn}

To understand the properties of our near-infrared spectroscopic sample in context with other DSFGs, and with star-forming galaxies in general, it is important to first understand the fraction of sources with active galactic nuclei (AGN) in our sample. We expect this to be moderate -- the largest sample of 870\,$\mu$m-selected DSFGs (also one of the main parent samples for KAOSS), the AS2UDS sample, contains an estimated 18\,$\pm$\,10 per cent sources with AGN components based on X-ray and photometric tests \citep{stach19}. Our rest-frame optical spectra allow us to also search for spectral indications of AGN. Therefore to provide a census of AGN in our sample we assess how many sources meet the following criteria:
\begin{itemize}
    \item \protect{flux ratio [N{\sc ii}]$\lambda$6583/H$\alpha$\,$>$\,0.8} \citep[e.g.,][]{wisnioski18};
    \item H$\alpha$ emission displays a broad component with a linewidth of FWHM $>$\,1000\,km\,s$^{-1}$ \citep[e.g.,][]{genzel14};
    \item presence of a luminous X-ray counterpart ($L_{\rm X}$\,$\geq$\,10$^{42}$\,erg\,s$^{-1}$) matched within 1$''$ \citep{civano16,luo17,franco18,kocevski18};
    \item Spitzer IRAC colours indicating an AGN according to the criteria of \cite{donley12}.
\end{itemize}
In total 15 of the sample of 27 sources fit one or more of these criteria, indicating an AGN fraction of 56\,$\pm$\,14 per cent. This is significantly higher than the range quoted for the AS2UDS sample in \cite{stach19}, but we expect a bias towards detecting AGN in our sample given the fact that such sources display stronger emission lines. Additionally, our estimate should be treated as an upper limit given that some sources may meet the criteria for other reasons, for example high [N{\sc ii}]/H$\alpha$ ratios may also arise from high metallicities \citep[e.g.,][]{allen08,kewley08}. Tellingly, only one source, AS2UDS0492.0, meets all four criteria.

We therefore separate potential AGN-host sources into two categories: those that are classified as hosting AGN based on their rest-frame optical spectra, and those that are classified as hosting AGN based on their X-ray and/or IRAC properties. In the former case we include the 10 sources (37\,$\pm$\,12 per cent of the sample) that have [N{\sc ii}]/H$\alpha$\,$>$\,0.8 or FWHM$_{{\rm H}\alpha}>$\,1000\,km\,s$^{-1}$. In the latter category we find 13 sources (48\,$\pm$\,13 per cent), including eight that are also in the spectral AGN sample. In all plots that follow we flag AGN-classified DSFGs with a star symbol, with the star-forming sources shown as circular points.

%
%
\section{Results and discussion}
\label{sec:results_discussion}

We have identified a sample of 27 DSFGs with spatially resolved emission line maps, yielding velocity and velocity dispersion maps, along with their physical properties from SED fitting. Additionally, we have identified and flagged which of the sources exhibit properties that suggest a significant contribution to the emission from an AGN. We now turn to deriving the rotational velocities and intrinsic velocity dispersions of these galaxies, and the ratio of these two quantities that has been proposed as a diagnostic of the level of rotational support. We study the variation of these with other important properties such as star-formation rate and stellar mass, before placing our sample within the context of the Tully-Fisher relation and estimating the dynamical masses.

%
%
\subsection{Rotation curve modelling}
\label{sec:rot_curve_modelling}

From our derived resolved velocity maps (see \S\ref{sec:vel_maps}) we extract rotation curves and velocity dispersion profiles. Rotation curves are extracted from the velocity field within a 0.5$''$ wide ($N$\,=\,5 pixels in the rebinned cube) pseudo-slit along the PA$_{\rm kin}$ from the velocity field, taking the median of the pixels across the slit. Velocity dispersion profiles are extracted from the velocity dispersion maps using the same slit, and uncertainties are extracted from the corresponding uncertainty maps. To ensure that all sources are extended enough to study kinematics, we measure the extent of the rotation curve for each galaxy and only retain sources where this extent is greater than twice the FWHM of the PSF (see \S\ref{sec:sample}).

The resultant rotation curves are shown in Fig.~\ref{fig:kaoss_rot_curves_1}. We note that the velocities plotted are centred using the results of the Freeman disk fitting (described below). We also overlay the centred rotation curves onto the PV diagrams in Fig.~\ref{fig:kaoss_pv_diagrams_major}, which generally trace the flux well. Some PV diagrams appear to be more noisy than others, however this is expected given the modest S/N and seeing-limited nature of our observations, as well as considering the intrinsically complex and optically faint nature of our sources. To test the robustness of our results against S/N effects, we identified a subset of eight sources with more complex PV diagrams which are less well captured by the rotation curve modelling and recalculated all of the key trends and median properties involving $v_{\rm rot}$, $v_{\rm rot}/\sigma_0$, $v_{\rm circ}$ and $M_{\rm dyn}$ after excluding these sources. We find that this does not change our conclusions within the uncertainties. The same is true if we exclude the lower quartile in S/N (i.e. sources with S/N$_{{\rm H}\alpha}$\,$<$\,14.7) from the sample.

In order to derive rotational velocities we fit the rotation curves with a model of the form \citep{freeman1970}:
\begin{equation} \label{eq:freeman_disc}
    (v(r)-v_{\rm off})^2 = \dfrac{(r-r_{\rm off})^2\pi G\mu_0}{h}(I_0K_0-I_1K_1),
\end{equation}
following \cite{harrison17} and \cite{tiley21}, where $v$ is the velocity in km\,s$^{-1}$, $r$ is the radial distance from the centre along the rotation axis in kpc, $v_{\rm off}$ is the velocity offset of the rotation curve from the nominal systemic redshift (derived from the integrated spectra), $r_{\rm off}$ is the spatial offset of the selected centroid from the spatial position of the systemic velocity on the rotation curve, $\mu_0$ is the peak mass surface density, $h$ is the disc scale radius and $I_nK_n$ are Bessel functions evaluated at 0.5$r/h$. We see in Fig.~\ref{fig:kaoss_rot_curves_1} that this model is a reasonably good fit to the data for 24 of the 27 sources (89\,$\pm$\,18 per cent), which we term the {\it disc-like} sample. The best-fit parameters $v_{\rm off}$ and $r_{\rm off}$ are tabulated in the online supplementary table, and are used to centre the rotation curves/velocity dispersion profiles in Fig.~\ref{fig:kaoss_rot_curves_1} and the PV diagrams in Fig.~\ref{fig:kaoss_pv_diagrams_major}. In some cases the velocity offsets are as large as $|v_{\rm off}|$\,$>$\,100\,km\,s$^{-1}$ and the spatial offsets can be over $|r_{\rm off}|$\,$>$\,4\,kpc. As we take the redshift measured from the integrated emission this suggests that there are significant asymmetries in the H$\alpha$ intensity which may be an indicator of turbulent structure in our sample. A similar argument can be made for the spatial offsets, as the initial centres are in most cases derived from the rest-frame optical continuum emission.

For the three sources which are poorly fit by the Freeman disc model we are unable to estimate a robust rotational velocity. Therefore, we attempt to assess whether there are any obvious intrinsic differences in the physical properties of these three sources with the 24 {\it disc-like} sources. Examining median values for the two subsamples, the {\it disc-like} and poorly fit samples have consistent stellar masses, $M_\ast$\,$=$\,(1.3\,$\pm$\,0.3)\,$\times$\,10$^{11}$\,M$_\odot$ and $M_\ast$\,$=$\,(3\,$\pm$\,4)\,$\times$\,10$^{11}$\,M$_\odot$, consistent star-formation rates, SFR\,$=$\,220\,$\pm$\,30\,M$_\odot$\,yr$^{-1}$ and SFR\,$=$\,160\,$\pm$\,60\,M$_\odot$\,yr$^{-1}$, and consistent dust extinctions, $A_V$\,$=$\,1.98\,$\pm$\,0.15 and $A_V$\,$=$\,2.6\,$\pm$\,0.5, respectively. We therefore find no significant difference between the two subsets in terms of their physical properties.

Additionally, we test for differences in the integrated S/N of the H$\alpha$ emission and $K$-band magnitudes of the two subsamples, respectively finding median S/N\,$=$\,19\,$\pm$\,4 and S/N\,$=$\,17\,$\pm$\,16, and $K$\,$=$\,21.4\,$\pm$\,0.2 and $K$\,$=$\,21.0\,$\pm$\,0.8. Therefore neither of these observed properties show any significant difference between the two subsamples. Given these findings, we elect to focus the remainder of our analysis on the 24 sources with robust rotational velocity measurements, and suggest that the three poorly fit systems are likely comparable to the {\it disc-like} sources but potentially with high pressure support, therefore omitting them does not significantly bias our conclusions.

\begin{figure*}
    \centering
    \includegraphics[width=0.49\linewidth]{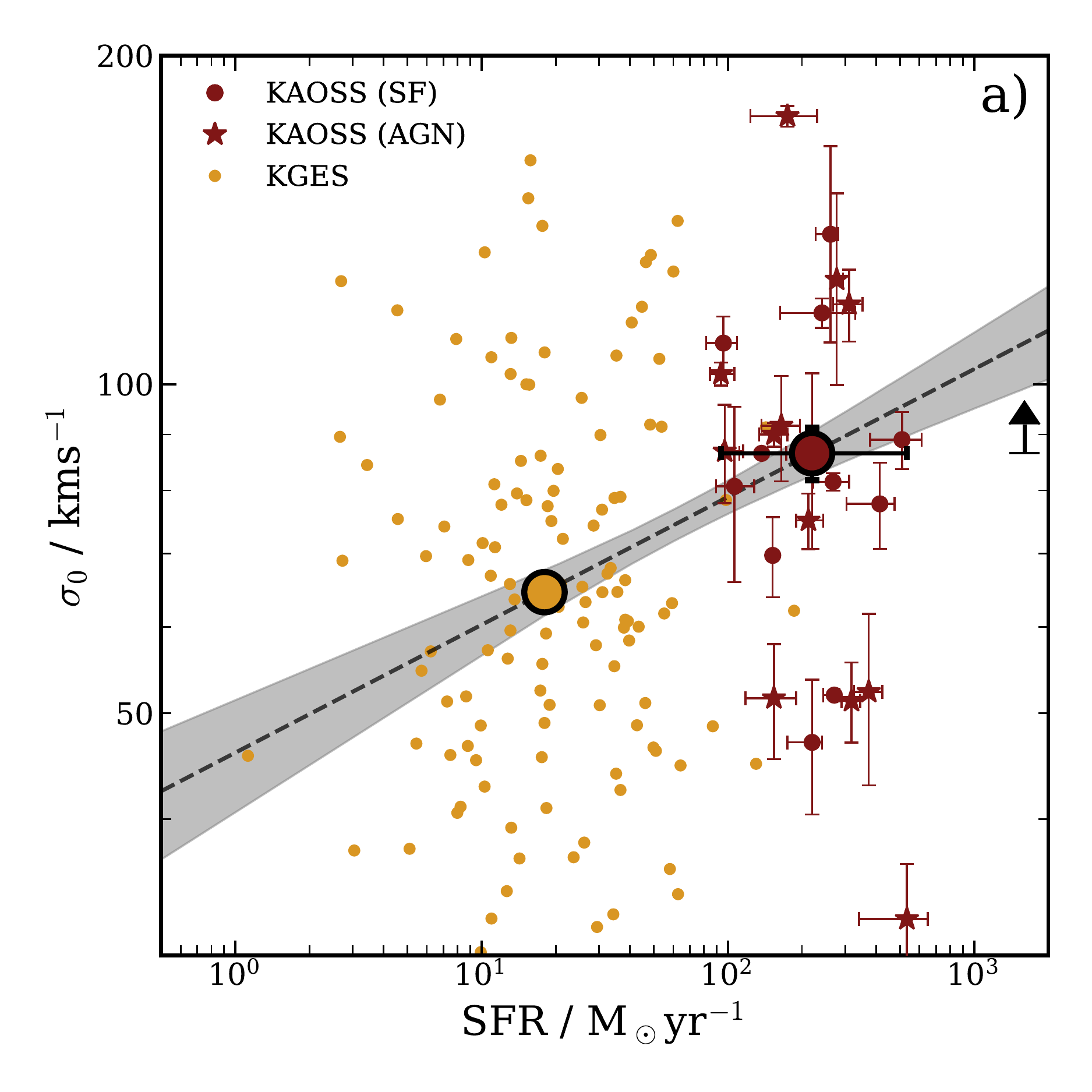}
    \includegraphics[width=0.49\linewidth]{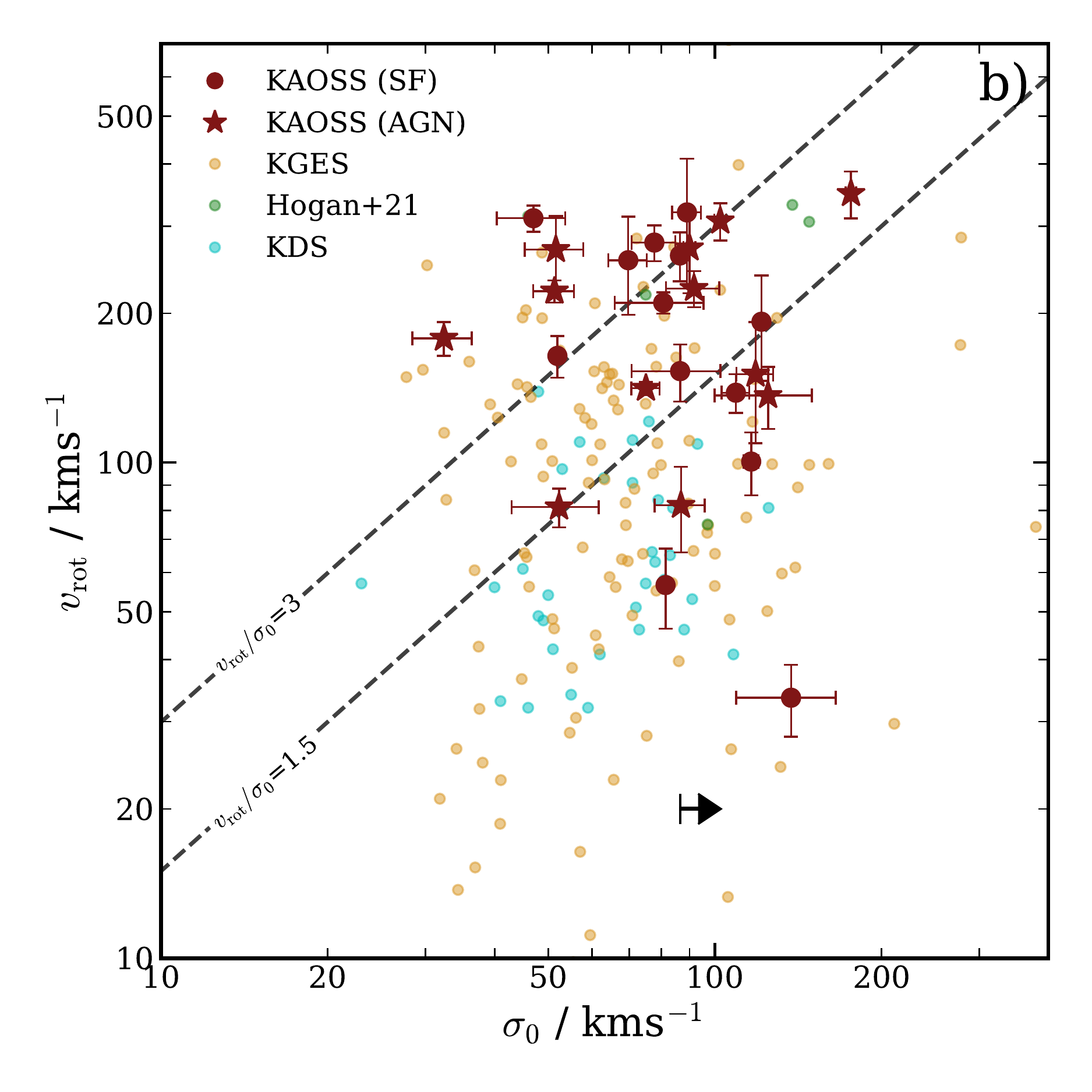}
    \caption{{\bf a)} Intrinsic velocity dispersion ($\sigma_0$) versus {\sc magphys}-derived star-formation rate for the KAOSS sample of DSFGs and KGES sample of typical star-forming galaxies at $z$\,$\sim$\,1.5. The KAOSS resolved sample has star-formation rates that are roughly an order of magnitude higher than the KGES sample, and we see a positive trend in $\sigma_0$ with star-formation rate, although the influence of the different sample selections on this trend is uncertain. We note that the ``error bar'' on the abscissa on the KAOSS median point represents the range in star-formation rate spanned by the bin, and this is also the case in subsequent plots (for the corresponding quantities plotted on the $x$-axis). {\bf b)} Rotational velocity $v_{\rm rot}$ versus intrinsic velocity dispersion $\sigma_0$ for the KAOSS {\it disc-like} sample along with the typical star-forming galaxies in KGES and KDS (at $z$\,$\sim$\,1.5 and $z$\,$\sim$\,3 respectively), and the Herschel-selected $z$\,$\sim$\,2.5 ULIRGS from \protect\cite{hogan21}. We indicate lines of constant $v_{\rm rot}/\sigma_0$\,$=$\,1.5 and 3, two values frequently used to assess if a system is rotationally supported. In general, the KAOSS DSFGs are more rapidly rotating, with higher velocity dispersions than the other samples, which results in only slightly higher $v_{\rm rot}/\sigma_0$ than those galaxies. This result is important as it shows that {\it both} $v_{\rm rot}$ and $\sigma_0$ are driving the variation of $v_{\rm rot}/\sigma_0$ in DSFGs with respect to other galaxy populations.}
    \label{fig:v_vs_sigma}
\end{figure*}

%
%
\subsection{Inclination-corrected rotational velocities}
\label{sec:v_true}

As a measure of the rotational velocity of each galaxy we evaluate $v_{\rm rot}$\,$=$\,$v_{2.2R_{\rm d}}$\,/\,$\sin{i}$, where $v_{2.2R_{\rm d}}$ is the observed velocity at 2.2\,$R_{\rm d}$\footnote{$R_{\rm d}$ is also convolved with $\sigma_{\rm PSF}\sim$\,2\,kpc.} according to the model fits and the factor of 1\,/\,$\sin{i}$ corrects for the observed inclination of the source. The inclination angle $i$ is measured from {\sc galfit} modelling (see \S\ref{sec:galfit}, and the online supplementary table).

We note that as some of the derived inclination angles are apparently small ($i$\,$<$\,20$^\circ$) we set a minimum inclination of $i$\,$=$\,20$^\circ$, to avoid significant extrapolations (given the simple models we are adopting). This only affects two sources, GDS0046.0 and AS2UDS0292.0 that have {\sc galfit}-derived inclinations of $i$\,$=$\,12$^\circ$\,$\pm$\,17$^\circ$ and $i$\,$=$\,14$^\circ$\,$\pm$\,15$^\circ$ respectively. Both are consistent with our chosen minimum inclination ($i$\,$=$\,20$^\circ$) within their large uncertainties. Additionally, the axis ratio measured from {\sc galfit} for AS2UDS0707.0 is less than $(b/a)_0$ (for which we adopted a value of 0.2, see \S\ref{sec:galfit}), resulting in an unphysical inclination angle; we also set a minimum inclination of $i$\,$=$\,20$^\circ$ for this source.

After applying inclination corrections we derive a median rotational velocity of $v_{\rm rot}$\,$=$\,190\,$\pm$\,40\,km\,s$^{-1}$ for the {\it disc-like} sources. For context, the more typical star-forming galaxies at $z$\,$\sim$\,1.5 in KGES have a median inclination-corrected velocity at 2.2\,$R_{\rm d}$ of 61\,$\pm$\,5\,km\,s$^{-1}$, and those at $z$\,$\sim$\,1 from KROSS have a median inclination-corrected velocity at 2.2\,$R_{\rm d}$ of 109\,$\pm$\,5\,km\,s$^{-1}$, both of which are significantly lower than the corresponding value for the KAOSS {\it disc-like} sample. We conclude that KAOSS DSFGs have much higher rotational velocities than less active (and apparently lower mass) galaxies that have been observed with KMOS.

%
%
\subsection{Observed velocity dispersions and beam-smearing corrections}
\label{sec:beam_smearing}

We now turn to measuring the velocity dispersions in our sources, which along with the rotation velocities will allow us to assess the level of rotational and pressure support in DSFGs. To measure the observed velocity dispersion, $\sigma_{\rm obs}$, we inspect the velocity dispersion profiles (Fig.~\ref{fig:kaoss_rot_curves_1}) and divide them into two groups, following \cite{johnson18}: first, where the velocity dispersion appears to have flattened in the outskirts, we measure $\sigma_{\rm obs}$ as the median of the three outer points (spanning 0.3$''$ or $\sim$\,2.5\,kpc) on both sides and take the lower value of the two sides. In all other cases we simply measure $\sigma_{\rm obs}$ as the median of the dispersion profile. As in \cite{johnson18} we label the sources ``O'' and ``M'' (see Fig.~\ref{fig:kaoss_rot_curves_1}) to indicate that the intrinsic velocity dispersion has been measured from the ``outskirts'' or ``median'', respectively. For the 27 resolved sources, we measure $\sigma_{\rm obs}$ from the outskirts in 16 cases, and from the median in the remaining 11 cases.

As noted in \S\ref{sec:vel_maps}, our estimates of the rotational velocity and velocity dispersion are affected by beam smearing, which results in an increased velocity dispersion near the centre of the galaxy. To correct for this effect and obtain estimates of the intrinsic velocity dispersion $\sigma_0$ we use the prescriptions of \cite{johnson18}, who derived correction factors from mock KMOS observations for the KROSS survey. For specific details on the corrections we direct the reader to \cite{johnson18}, but we note here that they are dependent on the size and rotational velocity of the galaxy. For the former we adopt the median $R_{\rm d}$ of our sample from the {\sc galfit} measurements for all sources, $R_{\rm d}$\,$=$\,0.40\,$\pm$\,0.05$''$, and for the latter we use the observed velocity at 2.2\,$R_{\rm d}$. Observed and intrinsic velocity dispersions, $\sigma_{\rm obs}$ and $\sigma_0$, are listed in the online supplementary table.

For the 11 sources in the median (``M'') subset the median beam-smearing correction is $\xi_{\sigma,{\rm M}}$\,$=$\,0.63\,$\pm$\,0.07, and for the 16 sources in the outskirts (``O'') subset the median beam-smearing correction is $\xi_{\sigma,{\rm O}}$\,$=$\,0.96\,$\pm$\,0.02. This demonstrates that the effects of beam smearing are much less severe in the outskirts of the galaxy. The value for the ``O'' subset is comparable to the corrections used for KROSS by \cite{johnson18} who found $\xi_{\sigma,{\rm O}}$\,$=$\,0.96$^{+0.02}_{-0.06}$ for outskirt $\sigma$ measurements, but they applied a much less significant correction than we do, $\xi_{\sigma,{\rm M}}$\,$=$\,0.8$^{+0.1}_{-0.3}$ for their median $\sigma$ measurements. 

Investigating the cause of this difference, the KAOSS sources are marginally larger than the lower-SFR KROSS sources, median $R_{\rm e}$\,$=$\,3.6\,$\pm$\,0.3\,kpc for KAOSS compared to a median of $R_{\rm e}$\,$=$\,2.9$_{-1.5}^{+1.8}$\,kpc \citep{harrison17} for the lower-redshift systems in KROSS (both surveys have comparable seeing). However, the KAOSS galaxies have much higher rotational velocities, with a median $v_{\rm rot}$\,$=$\,190\,$\pm$\,40\,km\,s$^{-1}$ compared to a median $v_{\rm rot}$\,$=$\,109\,$\pm$\,5\,km\,s$^{-1}$ from KROSS \citep{harrison17}. Therefore the KAOSS observations experience stronger beam smearing than those of KROSS due to the much larger rotational velocities of the galaxies.

We also apply beam-smearing corrections to the rotation velocities, $\xi_v$, following \cite{johnson18}, although these corrections are generally much smaller. Our sample has a median rotational velocity correction of $\xi_v$\,$=$\,1.06\,$\pm$\,0.01, increasing the median $v_{\rm rot}$ by $\sim$\,11\,km\,s$^{-1}$. This is also consistent with the corrections applied by \cite{johnson18} to the KROSS sample, who found a median $\xi_v$\,$=$\,1.07\,$\pm$\,0.03.

We caution that our use of the beam-smearing corrections from \cite{johnson18} are based on the assumption that the resolved KAOSS sources can be described by rotating discs, which is a simplistic assumption for DSFGs given that they may be kinematically more complex, as discussed earlier \citep[see also e.g.,][]{alaghband-zadeh12}. We therefore add vectors to all figures that show quantities derived using $\sigma_0$ to illustrate how the plotted values would change if we had applied no beam-smearing correction. The ``true'' intrinsic velocity dispersions are likely to lie somewhere between no correction and the full correction. As the corrections to $v_{\rm rot}$ are small ($\sim$\,5 per cent) we do not add similar vectors to plots including the rotational velocity.

\begin{figure*}
    \centering
    \includegraphics[width=0.49\linewidth]{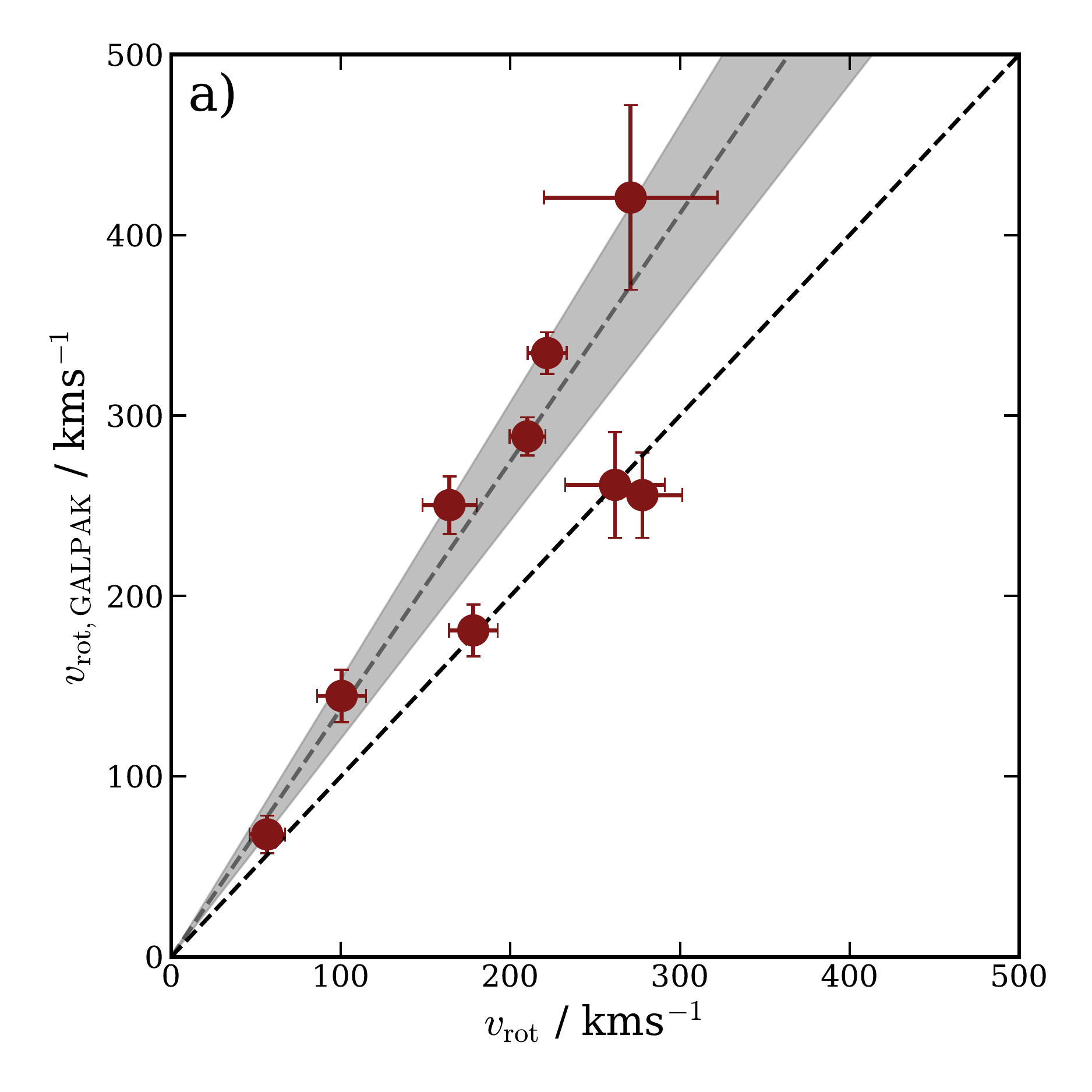}
    \includegraphics[width=0.49\linewidth]{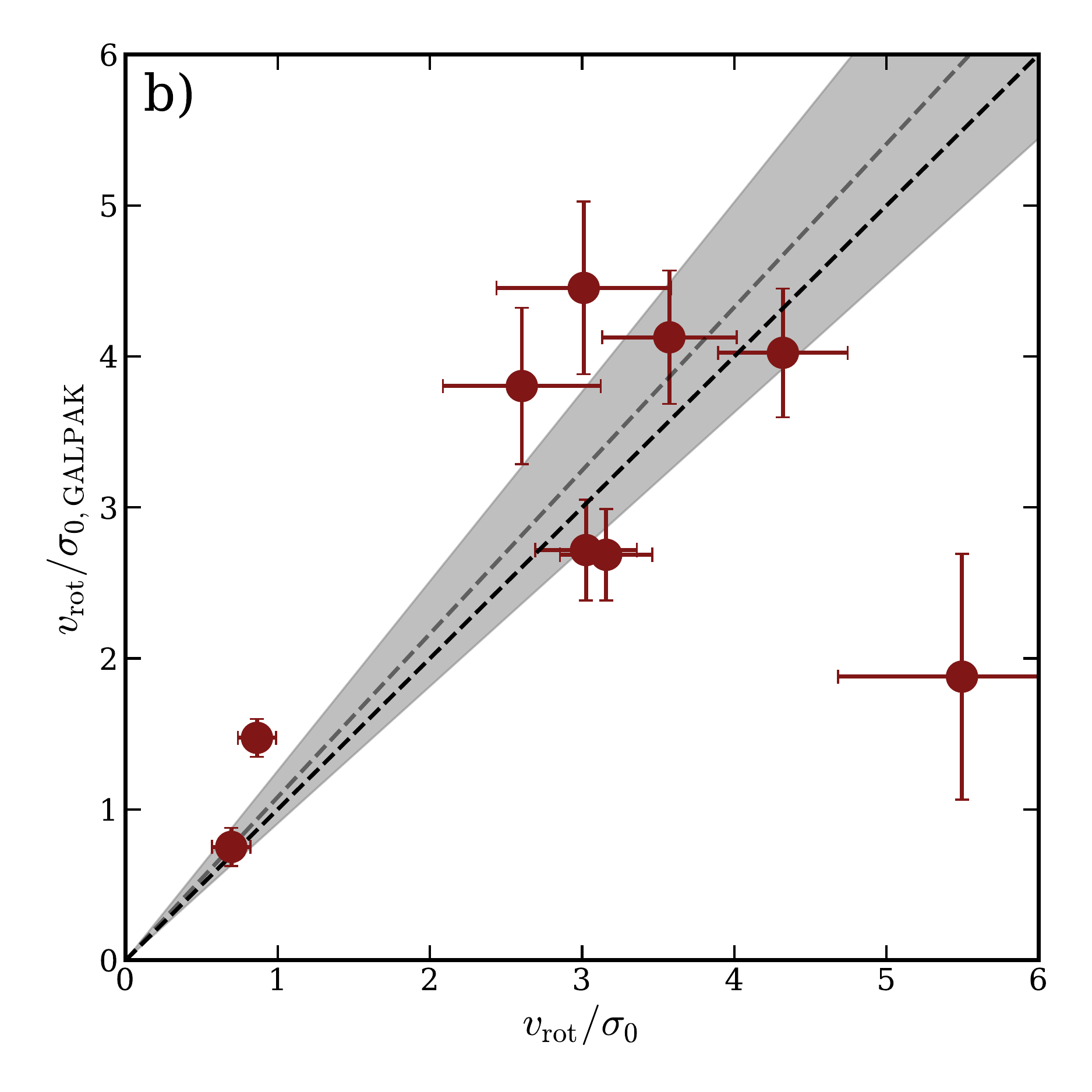}
    \caption{Comparison of the kinematic properties derived from our two-dimensional analysis (\S\ref{sec:rot_curve_modelling}) for the nine {\it disc-like} galaxies with values derived from {\sc galpak}$^{\rm 3D}$ modelling (\S\ref{sec:galpak}). In both cases the 1:1 line is dashed black and the grey dashed line shows the median ratio of the two sets of measurements with uncertainties indicated by the grey shaded region. {\bf a)} Rotational velocity at 2.2 times the disc radius. The two sets of values are consistent in most cases, with the exceptions having slightly lower values from {\sc galpak}$^{\rm 3D}$. The median ratio of $v_{\rm rot,GALPAK}/v_{\rm rot}$ is 1.4\,$\pm$\,0.2 indicating that where {\sc galpak}$^{\rm 3D}$ modelling can successfully converge on our data, it marginally overestimates rotational velocities measured from our 2-D analysis. {\bf b)} The ratio of $v_{\rm rot}/\sigma_0$. With the exception of AS2COS0048.1, where {\sc galpak}$^{\rm 3D}$ severely overestimates the velocity dispersion compared to our 2-D analysis, we see that the two sets of values are in reasonably good agreement within their uncertainties.}
    \label{fig:v_galpak_comparison}
\end{figure*}

%
%
\subsection{Intrinsic velocity dispersions}
\label{sec:vel_disps}

The intrinsic velocity dispersions $\sigma_0$ we have derived provide a measure of how turbulent our DSFGs are, and we can compare these values to those measured from other galaxy populations to determine the relative level of turbulence as a function of galaxy parameters such as rotational velocity and star-formation rate.

Our sample has a median $\sigma_{\rm obs}$\,$=$\,111\,$\pm$\,18\,km\,s$^{-1}$, and a median $\sigma_0$\,$=$\,87\,$\pm$\,6\,km\,s$^{-1}$. The \cite{hogan21} and \cite{alaghband-zadeh12} samples, which are similarly selected, have median intrinsic velocity dispersions of $\sigma_0$\,$=$100\,$\pm$\,20\,km\,s$^{-1}$ and $\sigma_0$\,$=$160\,$\pm$\,60\,km\,s$^{-1}$, respectively, both of which are comparable with the KAOSS sample. On the other hand the less active KGES sample has a median beam smearing-corrected velocity dispersion of $\sigma_0$\,$=$\,46\,$\pm$\,2\,km\,s$^{-1}$ \citep{tiley21}. Hence the dust-obscured and typically strongly star-forming KAOSS sources have systematically higher intrinsic velocity dispersions than the more typical KGES galaxies. However, this does not strictly imply that star-formation rate is the main cause of this difference.

Determining the origin of the turbulence is difficult, and we do not attempt to do this quantitatively here. Distinct kinematic components in our data would indicate ongoing interactions, which would likely produce high levels of turbulence from tidal flows between the two systems and internal torques, but our data do not generally reveal such components.
It is possible that AO-assisted SINFONI/ERIS or JWST/NIRSpec observations would uncover interactions (such as the features seen with JWST/NIRCam in some of the sources in Fig.~\ref{fig:nircam}), as they would more easily probe $\sim$\,1\,kpc scales. Alternatively, star formation itself may induce the turbulence, or misaligned cold accretion flows from the IGM, which could trigger the release of large amounts of gravitational energy \citep{genzel08}.

To quantify this further, in Fig.~\ref{fig:v_vs_sigma} we show $\sigma_0$ versus star-formation rate, as measured from {\sc magphys} SED fitting. The KAOSS DSFGs have an order of magnitude higher star-formation rates than the KGES sample, and we again see that KAOSS galaxies display higher velocity dispersions, whether corrected or uncorrected for beam smearing. We fit the median KAOSS and KGES points (with bootstrap uncertainties), finding a 4.3$\sigma$ correlation between $\sigma_0$ and SFR, which we view as modest given that the influence of the two different sample selections is uncertain. Therefore the high SFRs could produce the observed turbulence, but we interpret this result with caution given the different selections (and different masses) of the two samples.

\begin{figure*}
    \centering
    \includegraphics[width=0.49\linewidth]{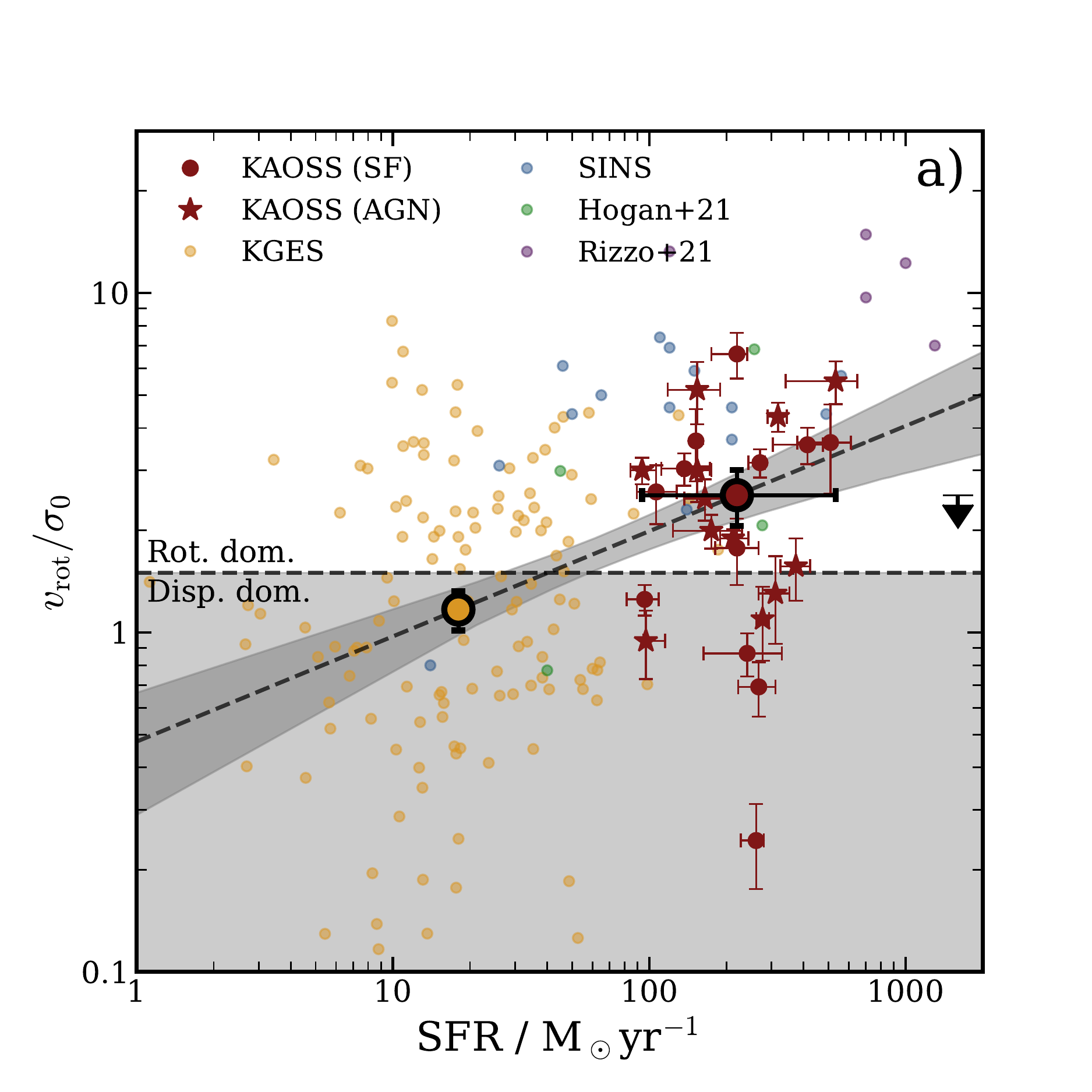}
    \includegraphics[width=0.49\linewidth]{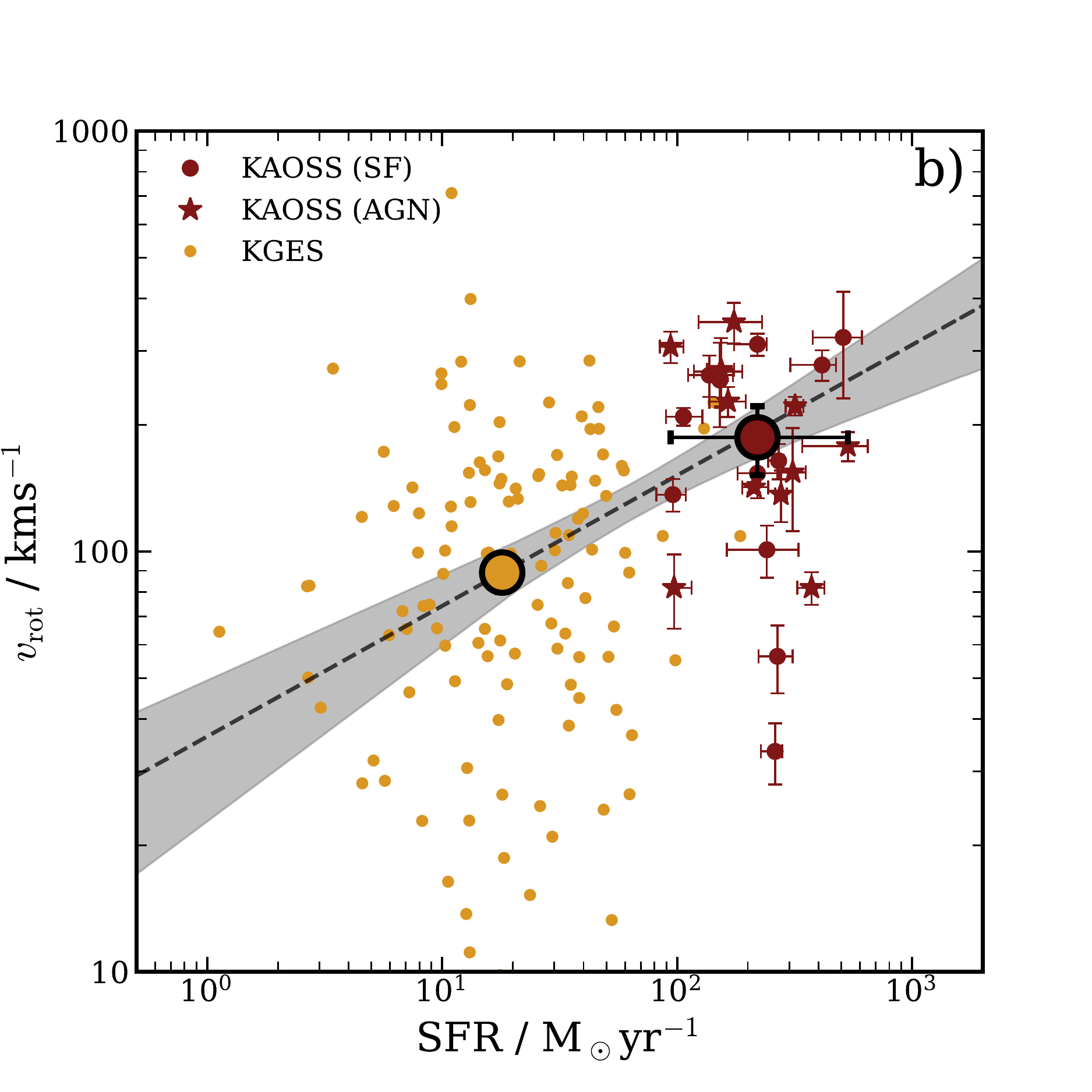}
    \caption{{\bf a)} Ratio of rotational velocity to velocity dispersion $v_{\rm rot}/\sigma_0$ versus star-formation rate. Also shown are $z$\,$\sim$\,1.5 typical star-forming galaxies from the KGES survey and $z$\,$\sim$\,1.3--2.6 galaxies from SINS, as well as $z$\,$\sim$\,2.5 ULIRGs from \protect\cite{hogan21} and $z$\,$\sim$\,4.5 DSFGs from \protect\cite{rizzo21}. We note however that the kinematics for the latter are based on [C{\sc ii}] observations, which are frequently systematically higher than those based on H$\alpha$ observations \citep[e.g.,][]{rizzo20,rizzo21,fraternali21,lelli21,roman-oliveira23}. We fit the binned KAOSS and KGES median points (large circles with black outline), finding a moderate (3.5$\sigma$ significance) correlation between $v_{\rm rot}/\sigma_0$ and star-formation rate (black dashed line with grey uncertainty region). The arrow shows how far the points would move down if we were to remove the beam-smearing corrections to the velocity dispersions. {\bf b)} Rotational velocity $v_{\rm rot}$ versus star-formation rate for the KAOSS and KGES samples. We measure a 3.8$\sigma$-significance correlation between $v_{\rm rot}$ and star-formation rate from the fit to the binned data (black dashed line with grey uncertainty region), which we suggest is simply driven by the ``main-sequence'' correlation between star-formation rate and stellar mass, leading to higher star-formation rates in galaxies with greater stellar masses and therefore faster rotation velocities.}
    \label{fig:v_s_v_sfr}
\end{figure*}

%
%
\subsection{{\sc galpak}$^{\rm 3D}$ analysis}
\label{sec:galpak}

Given the diversity of properties within the KAOSS sample, so far in \S\ref{sec:results_discussion} we have adopted a simple approach for estimating rotational velocities and velocity dispersions, including in our approach to beam-smearing corrections. However, there are some concerns about such statistical a-posteriori corrections, particularly when applied to modestly sized samples such as that presented here. For example, \cite{diteodoro15} show that beam smearing increases the observed velocity dispersion in the outskirts of the nearby galaxy NGC2403 by a factor of $\sim$\,3, much larger than the corrections we apply to sources with their dispersions estimated from the outskirts of their $\sigma_{\rm obs}$ profiles. 3-D fitting routines have been developed to attempt to circumvent these issues and model the dynamics of high-redshift galaxies, which include forward modelling to estimate intrinsic velocity dispersions.

To test our intrinsic velocity dispersions and rotational velocities against those derived from more sophisticated 3-D fitting routines, we perform an independent analysis of the KMOS data cubes using the {\sc galpak}$^{\rm 3D}$ code \citep{bouche15}. With {\sc galpak}$^{\rm 3D}$ we adopt a parametric model, modelling the light profile of our galaxies as an exponential disc with a scale height of 0.15\,$R_{\rm e}$, and adopting a Freeman disc to describe the kinematics for consistency with our previous 2-D analysis (see \S\ref{sec:rot_curve_modelling}). Our analysis follows \cite{puglisi23} and for further details we direct the reader to that work.

As {\sc galpak}$^{\rm 3D}$ is not designed to handle complex kinematics we limit this analysis to a subset of DSFGs that are selected from the {\it disc-like} sample with the highest S/N in the integrated spectra. This results in a sample of ten DSFGs that have S/N\,$\geq$\,15 in the integrated H$\alpha$ emission. This choice is made to ensure that the modelling can converge successfully. Additionally, we fix the centroid, inclination and position angle to the values measured from our photometry (see \S\ref{sec:rotation_axes}), rather than allowing these to vary as free parameters. The fitting procedure was successful for nine of the ten sources, the velocity and velocity dispersions maps for which are shown in Figs.~\ref{fig:galpak_fits_vel} and \ref{fig:galpak_fits_sig}, respectively. We see from Fig.~\ref{fig:galpak_fits_vel} that in the majority of cases GALPAK$^{\rm 3D}$ reproduces the structure of our empirically derived velocity maps (see \S\ref{sec:vel_maps}), however the model fails to capture some of the more complex details, such as GDS0001.0 and GDS0023.0. On the other hand, from Fig.~\ref{fig:galpak_fits_sig} we see that in all cases GALPAK$^{\rm 3D}$ does not recover the structure of the velocity dispersion maps from \S\ref{sec:vel_maps}, and so we suggest that seeing-limited velocity dispersion maps of the ionised gas component for optically faint dusty galaxies represent a challenge for forward-modelling tools such as GALPAK$^{\rm 3D}$.

For the nine sources that were successfully modelled we derive $v_{\rm rot}$ and $\sigma_0$ using {\sc galpak}$^{\rm 3D}$ and compare these to our original measurements. We show the {\sc galpak}$^{\rm 3D}$ best-fit observed velocity and velocity dispersion profiles in Fig.~\ref{fig:kaoss_rot_curves_1}, with values reported in the online supplementary table. As with the maps shown in Fig.~\ref{fig:galpak_fits_vel}, we generally see that {\sc galpak}$^{\rm 3D}$ models the velocity profiles well, but is often unable to fit the velocity dispersion profile appropriately (Fig.~\ref{fig:galpak_fits_sig}).

In Fig.~\ref{fig:v_galpak_comparison} we compare the values of $v_{\rm rot}$ and $v_{\rm rot}/\sigma_0$ measured from {\sc galpak}$^{\rm 3D}$ with those measured from our 2-D analysis (\S\ref{sec:v_true} and \S\ref{sec:rot_support}). This shows that, as we can see qualitatively in Fig.~\ref{fig:kaoss_rot_curves_1}, $v_{\rm rot}$ is reasonably well-captured by {\sc galpak}$^{\rm 3D}$, and we measure a median value of $v_{\rm rot,GALPAK}/v_{\rm rot}$\,$=$\,1.4\,$\pm$\,0.2. Therefore, {\sc galpak}$^{\rm 3D}$ derives a slightly higher rotational velocity compared to our empirical measurements. This may imply that the required beam-smearing corrections are higher than those we estimated in \S\ref{sec:beam_smearing}. On the other hand, {\sc galpak}$^{\rm 3D}$ reproduces our measurements of $v_{\rm rot}/\sigma_0$ reasonably well, and we find a median $(v_{\rm rot}/\sigma_{0{\rm ,GALPAK}})/(v_{\rm rot}/\sigma_0)$\,$=$\,1.1\,$\pm$\,0.2. The outlier here is AS2COS0048.1, which has $v_{\rm rot}/\sigma_0$\,$\simeq$\,5.5, the highest in the {\sc galpak}$^{\rm 3D}$ sample. {\sc galpak}$^{\rm 3D}$ overestimates the intrinsic velocity dispersion by a factor of $\sim$\,3, and therefore $v_{\rm rot}/\sigma_0$ is significantly underestimated.

\begin{figure*}
    \centering
    \includegraphics[width=0.49\linewidth]{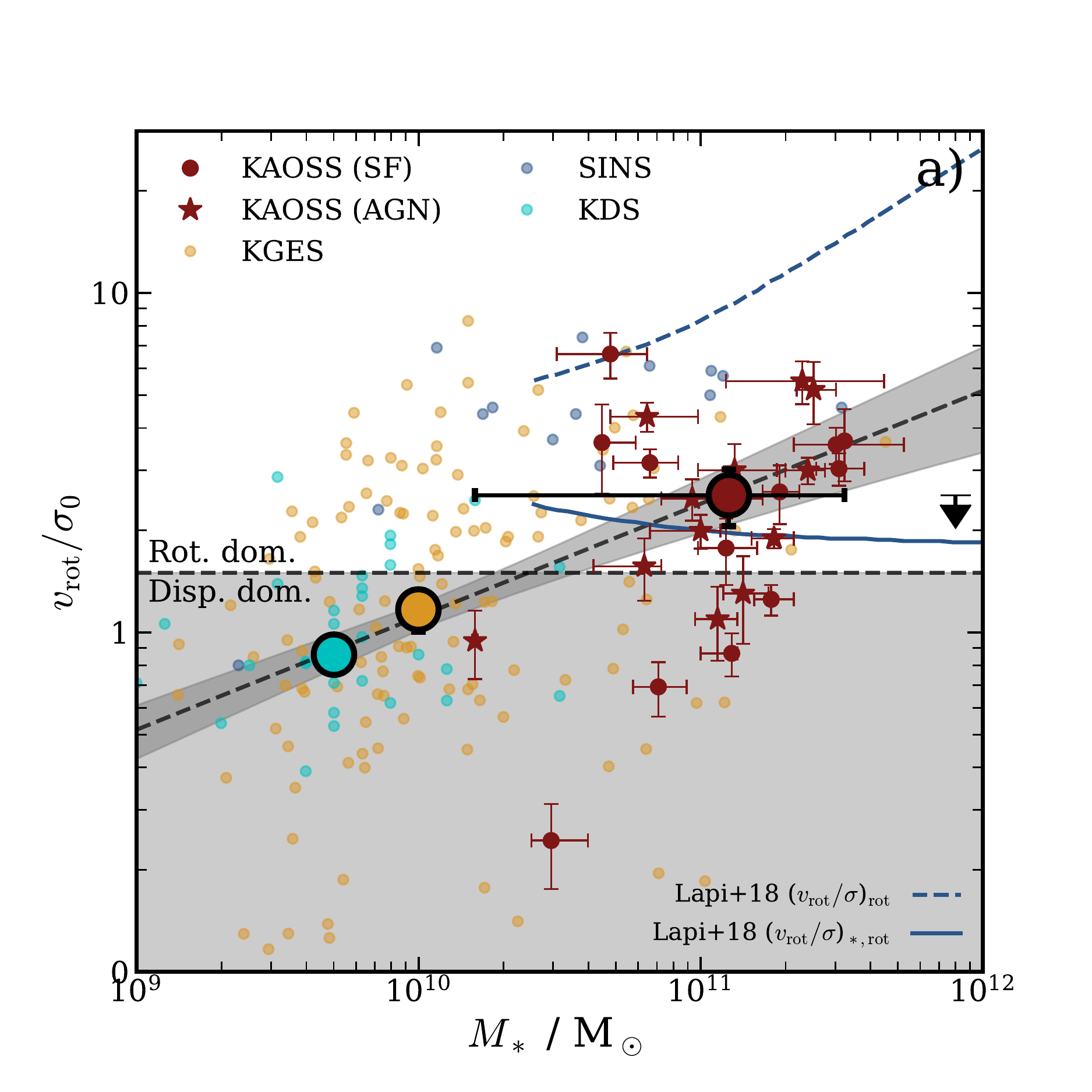}
    \includegraphics[width=0.49\linewidth]{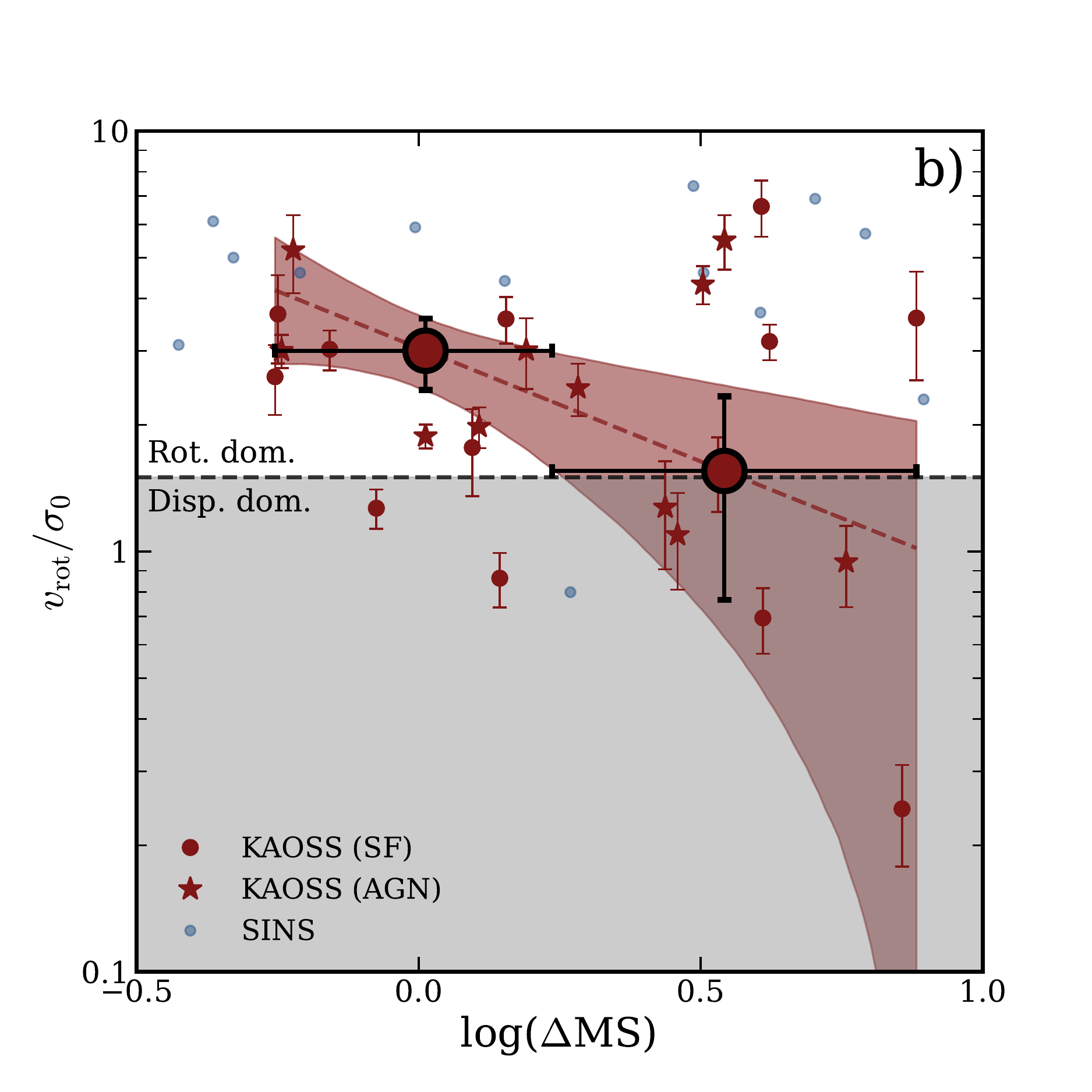}
    \caption{{\bf a)} Ratio of rotational velocity to velocity dispersion ($v_{\rm rot}/\sigma_0$) versus stellar mass showing the same galaxy samples as in the Fig.~\ref{fig:v_s_v_sfr}, with the addition of $z$\,$\sim$\,3.5 typical star-forming galaxies from the KMOS Deep Survey \protect\citep{turner17}. We include two tracks from simulations of early-type galaxy progenitors \protect\citep{lapi18}, with $v_{\rm rot}/\sigma_0$ measured at the centrifugal size of the gas (dashed) and stellar (solid) components. Our data are in reasonable agreement with the latter and are therefore consistent with the predictions of $v_{\rm rot}/\sigma_0$ for early-type progenitors. {\bf b)} Ratio of rotational velocity to velocity dispersion ($v_{\rm rot}/\sigma_0$) compared to offset from the main sequence ($\Delta$MS) for KAOSS and SINS star-forming galaxies. We fit our binned data (red dashed line and shaded region), which shows a weak trend at most between the two properties, suggesting that ``starburst''-like sources do not show a stronger contribution from pressure support than regular disc-like sources.}
    \label{fig:v_s_mstar}
\end{figure*}

%
%
\subsection{Rotational support}
\label{sec:rot_support}

Having determined that KAOSS DSFGs are apparently turbulent and massive sources, we now assess whether turbulence is the dominant source of motion. One of the simplest methods of doing this is to calculate the ratio of rotation velocity to intrinsic velocity dispersion $v_{\rm rot}/\sigma_0$ \citep[e.g.,][]{weiner06,wisnioski15}: if a galaxy has a much higher rotation velocity than its velocity dispersion then it is considered to be ``rotationally supported''. Alternatively a galaxy that appears to be predominantly pressure supported may be displaying inclination or projection effects, and/or may be interacting or merging.

Before deriving $v_{\rm rot}/\sigma_0$ we show the intrinsic velocity dispersion $\sigma_0$ as a function of the rotational velocity $v_{\rm rot}$ for the resolved KAOSS sources in Fig.~\ref{fig:v_vs_sigma}, alongside galaxies from the KGES sample and \cite{hogan21}. This demonstrates the elevated rotational velocities and velocity dispersions of the KAOSS sources compared to the KGES sample that were discussed in \S\ref{sec:rot_curve_modelling} and \S\ref{sec:vel_disps}.

Among the sources in our sample with robust $v_{\rm rot}$ measurements, 17 of the 24 sources (71\,$\pm$\,17 per cent) fit the criterion for rotationally dominated sources of $v_{\rm rot}/\sigma_0$\,$>$\,1.5, dropping to 11 out of 24 (46\,$\pm$\,14 per cent) if we instead adopt the criterion of $v_{\rm rot}/\sigma_0$\,$>$\,3. The median value of the {\it disc-like} sample is $v_{\rm rot}/\sigma_0$\,$=$\,2.5\,$\pm$\,0.5. This is consistent with the typical star-forming galaxies at $z$\,$\sim$\,1 from the KROSS sample \citep{stott16}, which has an average $v_{\rm rot}/\sigma_0$\,$=$\,2.2\,$\pm$\,1.4 (where the uncertainty is the standard deviation of the distribution), and the KGES sample which has a median $v_{\rm rot}/\sigma_0$\,$=$\,1.6\,$\pm$\,0.1. If we make the reasonable assumption of pressure support in the three sources that are not well fit by a Freeman disc model, 17 of the 27 sources, or 63\,$\pm$\,15 per cent of the DSFGs are rotationally supported ($v_{\rm rot}/\sigma_0$\,$>$\,1.5). We suggest that just over half of the resolved KAOSS DSFGs are likely to be rotationally supported systems with a not insignificant contribution from pressure support.

To further quantify this, we consider the circular velocity $v_{\rm circ}$:
\begin{equation}\label{eq:v_circ}
    v_{\rm circ} = \sqrt{ \, v_{{\rm 2.2}R_{\rm d}}^2 + \sigma_0^2\left(\dfrac{R}{R_{\rm d}}\right)},
\end{equation}
where the $\sigma_0^2$ term is a correction for turbulent motions (commonly referred to as the asymmetric drift correction), which contribute to the dynamical support of the system thus reducing the necessary rotational support for a stable orbit. A more general definition of the circular velocity is $v_{\rm circ}^2 = -R\nabla\phi$, where $\phi$ is the total gravitational potential. Eq.~\ref{eq:v_circ} applies under the assumption of an exponential disc profile, an isotropic velocity dispersion tensor, a radially constant velocity dispersion profile and a radially constant disc thickness \citep{lelli14}. As before, we estimate $v_{\rm circ}$ at $R$\,$=$\,2.2\,$R_{\rm d}$. We note that the coefficient of the asymmetric drift term is dependent on the assumed structure of the disc and the velocity dispersion profile, and typically ranges between 1 and 2. This is discussed in more detail in \cite{bouche22}, and for simplicity we adopt a value of unity in our analysis.

For the KAOSS sample we determine that the median $v_{\rm circ}$\,$=$\,230\,$\pm$\,20\,km\,s$^{-1}$. This $v_{\rm circ}$ is consistent with those seen in the most massive spiral galaxies at $z$\,$\sim$\,0 \citep{ogle16,diteodoro21}, however the DSFGs are seen $\sim$\,10\,Gyrs ago, have a higher space density than these present day ``super spirals'' \citep[e.g.,][]{dudzeviciute20}, and are expected to reside in denser environments at $z$\,$\sim$\,0. Adopting a coefficient of 2 on the asymmetric drift term gives a $\sim$\,25 per cent higher value of $v_{\rm circ}$\,$=$\,290\,$\pm$\,20\,km\,s$^{-1}$. For our chosen value of the coefficient (i.e. unity), we find the median contribution of $v_{\rm rot}$ and $\sigma_0$ to $v_c$ is 75 per cent and 25 per cent on average, respectively. Changing the coefficient from 1 to 2 results in corresponding percentages of 60 per cent and 40 per cent. In what follows, we note the influence of changing the coefficient on our conclusions where relevant.

Our DSFGs are among the more actively star-forming systems at $z$\,$\sim$\,2, with a median SFR of 210\,$\pm$\,30\,M$_\odot$yr$^{-1}$. As such, we are interested in understanding the implications of this fact for the kinematics of the sources. We showed in \S\ref{sec:vel_disps} that KAOSS sources appear to have higher velocity dispersions than the $z$\,$\sim$\,1.5 ``main sequence'' KGES galaxies, as traced by their H$\alpha$ emission. We are also interested in whether $v_{\rm rot}/\sigma_0$ varies similarly with SFR, and we show these two quantities in Fig.~\ref{fig:v_s_v_sfr}. To search for a trend between $v_{\rm rot}/\sigma_0$ and SFR we bin the KAOSS and KGES sources in SFR, but we see little evidence for more highly star-forming sources being significantly more or less rotation dominated, and fitting the binned points reveals a positive trend that is only marginally significant (given the uncertain influence of the differing sample selections) at the 3.5$\sigma$ level. To test the driver of this relation we study the $v_{\rm rot}$ versus SFR (in Fig.~\ref{fig:v_s_v_sfr}) and $\sigma_0$ versus SFR (in Fig.~\ref{fig:v_vs_sigma}, discussed in \S\ref{sec:vel_disps}) trends. We find the $v_{\rm rot}$--SFR and $\sigma_0$--SFR relations to have 3.8$\sigma$ and 4.3$\sigma$ positive correlations (before consideration of selection effects), respectively when considering the KAOSS and KGES binned medians.

The $v_{\rm rot}$--SFR correlation likely reflects the so-called ``main sequence'' trend whereby galaxies with larger stellar masses have higher star-formation rates \citep[e.g.,][]{brinchmann04,elbaz07,noeske07a,whitaker12,schreiber15}, and as a result of their higher stellar masses they also exhibit higher rotational velocities. In conclusion, we have little evidence to suggest that KAOSS DSFGs are more or less rotation dominated (as measured by $v_{\rm rot}/\sigma_0$) than less active SFGs, and they may simply be scaled-up versions of such sources, which are more massive but with similar relative levels of rotational and pressure support.

We also test for a correlation between $v_{\rm rot}/\sigma_0$ and stellar mass, which is shown in Fig.~\ref{fig:v_s_mstar}, along with more typical star-forming galaxies at lower redshifts from KDS and KGES. We fit the binned medians of these samples, finding a 4.7$\sigma$ positive correlation between the two quantities, suggesting that galaxies with higher stellar masses are more rotation dominated as expected. In Fig.~\ref{fig:v_s_mstar} we also include theoretical predictions by \cite{lapi18} for the descendants of the local early-type galaxy (ETG) population. These predictions include $v_{\rm rot}/\sigma_0$ measured at several different radii, and we include here only the values at the gas and stellar centrifugal sizes. Our data are consistent with their $v_{\rm rot}/\sigma_0$ values measured at the stellar centrifugal size.

One of the predictions from the observed ``main sequence'' is that galaxies within its spread are secularly evolving, whereas sources significantly above the main-sequence SFRs experience a different mechanism driving star formation, such as mergers or interactions. Therefore, we may expect sources above the main sequence to have lower $v_{\rm rot}/\sigma_0$. To test this, we show in Fig.~\ref{fig:v_s_mstar} $v_{\rm rot}/\sigma_0$ versus $\Delta$MS, i.e. the specific star-formation rate (sSFR) normalised by the main-sequence sSFR (for its measured stellar mass and redshift). Galaxies with higher $\Delta$MS are more ``starburst''-like. We adopt the \cite{speagle14} prescription of the main sequence. Fig.~\ref{fig:v_s_mstar} shows $v_{\rm rot}/\sigma_0$ versus log($\Delta$MS) for KAOSS and SINS galaxies (where $\Delta$MS is evaluated at the individual redshift of each source), and we divide our data into two bins which show no significant correlation with $v_{\rm rot}/\sigma_0$. We therefore see no evidence to suggest that the main-sequence-normalised sSFRs of our sources correlate with rotational support.

\begin{figure}
    \centering
    \includegraphics[width=\linewidth]{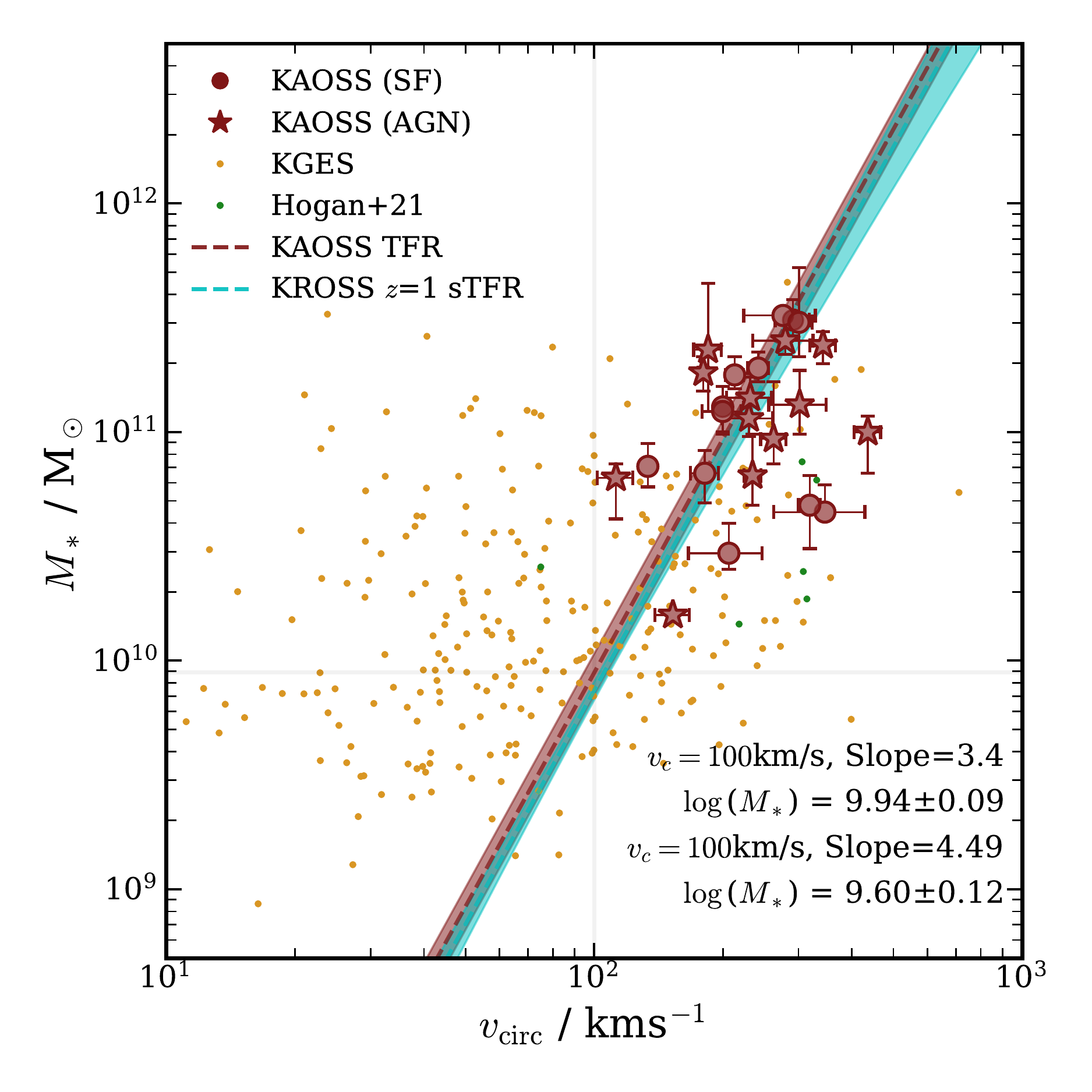}
    \caption{Stellar Tully-Fisher relation (sTFR): $v_{\rm circ}$ versus stellar mass for the KAOSS sample of DSFGs and KGES sample of typical star-forming galaxies at $z$\,$\sim$\,1.5, along with $z$\,$\sim$\,2.5 ULIRGs from \protect\cite{hogan21}. The sTFR derived by \protect\cite{tiley19} for $z$\,$\sim$\,1 typical galaxies from KROSS is shown by the cyan dashed line and shaded region. We fit for the normalisation of the Tully-Fisher relation, fixing the slope to be $a$\,$=$\,3.4 \protect\citep{tiley19} and finding a best-fit normalisation of $\log_{10}(M_\ast/{\rm M}_\odot)$\,$=$\,9.94\,$\pm$\,0.09 at $v_{\rm circ}$\,$=$\,100\,km\,s$^{-1}$, consistent with the value found by \protect\cite{tiley19} for $z$\,$=$\,0 SAMI galaxies, $\log_{10}(M_\ast/{\rm M}_\odot)$\,$=$\,9.87\,$\pm$\,0.04. }
    \label{fig:tully_fisher}
\end{figure}

\begin{figure*}
    \centering
    \includegraphics[width=0.49\linewidth]{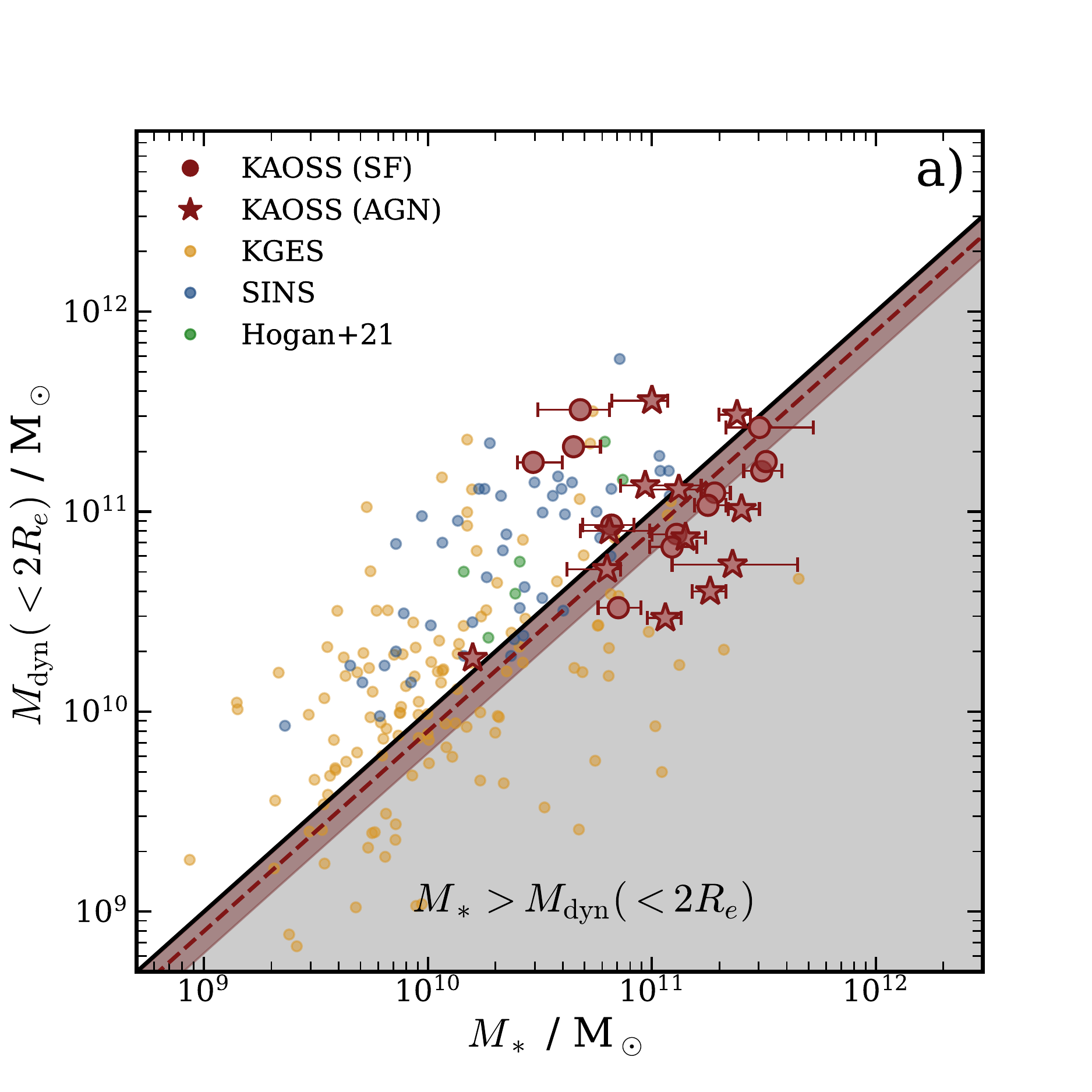}
    \includegraphics[width=0.49\linewidth]{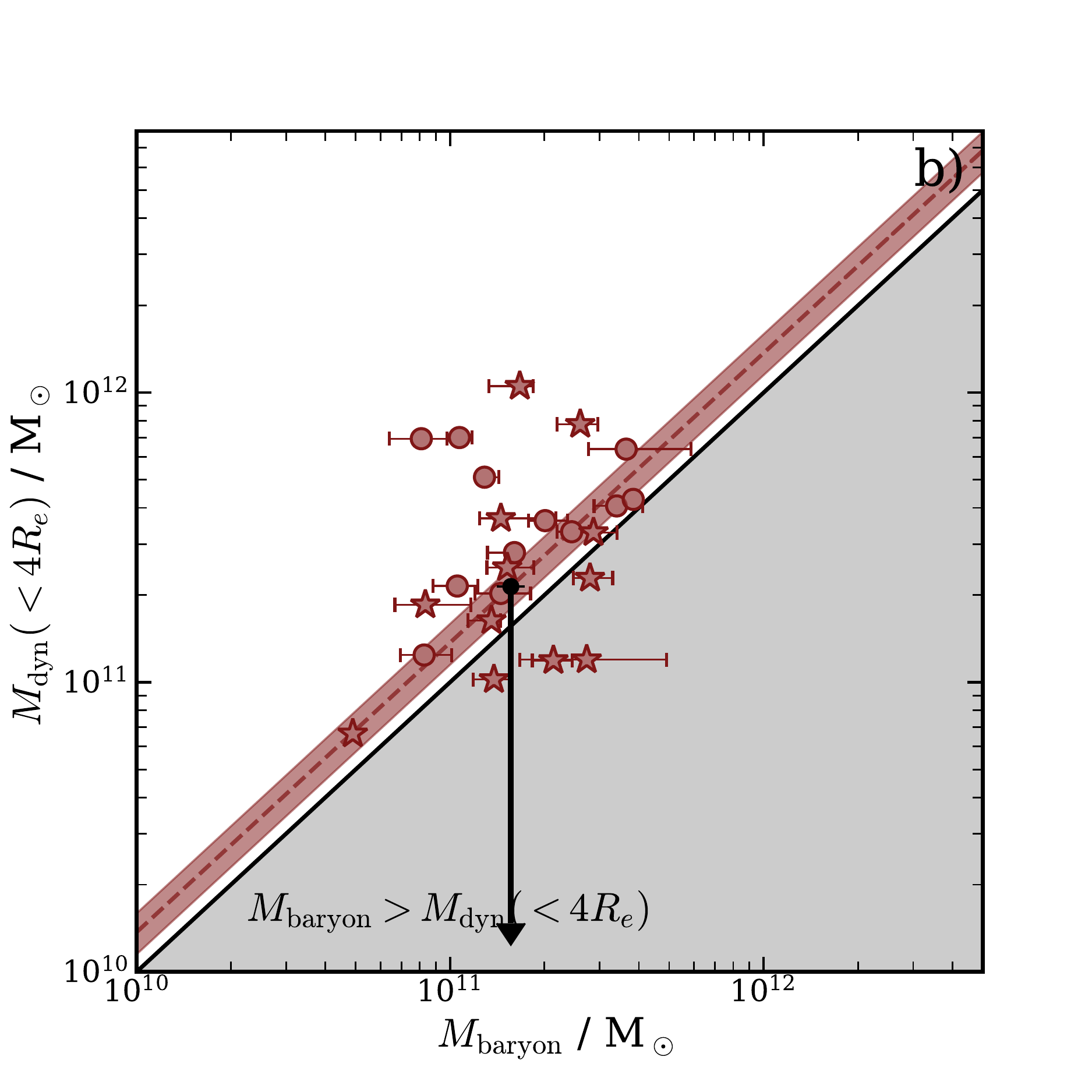}
    \caption{{\bf a)} Dynamical masses of the resolved KAOSS sample, estimated from the H$\alpha$ kinematics within 2$R_{\rm e}$, plotted against stellar masses. Plotted as a comparison are $z$\,$\sim$\,1.5 KGES main-sequence galaxies \protect\citep{tiley21}, $z$\,$\sim$\,2 SINS galaxies \protect\citep{forster-schreiber09} and $z$\,$\sim$\,2.5 ULIRGs from \protect\cite{hogan21}. The black solid line indicates the 1:1 relation, and therefore points below this have stellar masses that are greater than their dynamical masses within 2$R_{\rm e}$. The red dashed line indicates a fit to the KAOSS data points, with uncertainties shown by the shaded region. {\bf b)} The same as {\bf a)} except here dynamical masses are estimated within 4$R_{\rm e}$, and now plotted against baryonic mass, adding a gas component to the stellar mass by converting the {\sc magphys} dust masses to gas masses using a gas-to-dust ratio of $\delta_{\rm gdr}$\,$=$\,65 \protect\citep[][]{birkin21}. The black solid line and red dashed line are the same as in {\bf a)}. The vertical arrow indicates how far down the points would move if we were to adopt an aperture of 2$R_{\rm e}$, in which case the baryonic masses are typically greater than the dynamical masses. We therefore suggest that much of the baryonic matter (in particular the cold molecular gas) is situated at radii larger than the stars in DSFGs.}
    \label{fig:dynamical_masses}
\end{figure*}

%
%
\subsection{Tully-Fisher relation}
\label{sec:TFR}

The Tully--Fisher relation \citep[TFR;][]{tully1977} relates the stellar or baryonic matter content of a galaxy to its total mass, including dark matter. Our sample, which is one of the largest with estimates of kinematic information for DSFGs, allows us to probe the TFR for this massive galaxy population at $z$\,$\sim$\,1.5--2.5. The TFR uses the circular velocity of the interstellar medium as a proxy for the potential of the galaxy halo.

Fig.~\ref{fig:tully_fisher} shows the estimated circular velocity $v_{\rm circ}$ for the {\it disc-like} KAOSS resolved subset versus their stellar masses as estimated from {\sc magphys} SED fitting (see \S\ref{sec:magphys}). We also include similar measurements for KGES galaxies \citep{tiley21} and Herschel-selected $z$\,$\sim$\,2.5 ULIRGs from \cite{hogan21}. In order to quantify the stellar Tully--Fisher relation (sTFR) for our $z$\,$\sim$\,2 sample, and to test for any evolution against samples at lower redshifts, we fit to our data points the model $\log_{10}(v_{\rm circ})$\,$=$\,$a \log_{10}(M_\ast)+b$ where $a$ and $b$ are constant parameters, using an orthogonal distance regression (ODR) method that takes both the errors in $M_\ast$ and $v_{\rm circ}$ into account.

Following \cite{tiley19} we measure the value of $\log_{10}(M_\ast)$ at $v_{\rm circ}$\,$=$\,100\,km\,s$^{-1}$ and find $\log_{10}\left(M_\ast/{\rm M}_\odot\right)_{v_{\rm circ}=100}$\,$=$\,9.94\,$\pm$\,0.09 from the fit to our data. This is not significantly higher than the values measured by \cite{tiley19} for both the $z$\,$\sim$\,1 KROSS (9.89\,$\pm$\,0.04) and $z$\,$\sim$\,0 SAMI (9.87\,$\pm$\,0.04) samples, indicating little or no evolution in the sTFR between $z$\,$\sim$\,2 and $z$\,$\sim$\,0. In comparison with other studies, \cite{bell01} find $\log_{10}(M_\ast)$\,$=$\,9.5 at $v_{\rm max}$\,$=$\,100\,km\,s$^{-1}$ at $z$\,$=$\,0, and \cite{conselice05} find corresponding values of 9.43\,$\pm$\,0.12 and 9.39\,$\pm$\,0.13 and $z$\,$<$\,0.7 and $z$\,$>$\,0.7, respectively. Both of these studies adopt a slope of 4.49. If we fix our slope to this value, we derive $\log_{10}(M_\ast)$\,$=$\,9.60\,$\pm$\,0.12 at $v_{\rm max}$\,$=$\,100\,km\,s$^{-1}$ (this is noted in the legend of Fig.~\ref{fig:tully_fisher}, but the fit itself is not shown). As with our comparison to the \cite{tiley19} samples, we do not see significant evidence for a change in the normalisation of the sTFR between $z$\,$\sim$\,2 and the present day.

%
%
\subsection{Dynamical masses}
\label{sec:mdyn}

The dynamical mass, i.e.\ the total matter content contributing to the motions of the material within the galaxy, is another important quantity that is not yet well-measured for many DSFGs. We estimate dynamical masses within twice the effective radius, 2$R_{\rm e}$ (which is typically $\sim$\,7\,kpc) for our sample according to:
\begin{equation}
    M_{\rm dyn}(<2R_{\rm e}) = \dfrac{2R_{\rm e} v_{\rm circ}^2}{G}
\label{eq:mdyn}
\end{equation}
following \cite{burkert10}, with $R$\,$=$\,2$R_{\rm e}$. Here we also measure $v_{\rm circ}$ at 2$R_{\rm e}$. In Fig.~\ref{fig:dynamical_masses} we show the dynamical mass estimates for our sample plotted against their stellar masses. Dynamical and baryonic masses are tabulated in the online supplementary table. We caveat here that Eq.~\ref{eq:mdyn} is uncertain for our sources given the requirement of spherical symmetry and the assumptions needed to derive their intrinsic velocity dispersions. We apply Eq.~\ref{eq:mdyn} for comparison with the literature \citep[see e.g.,][]{hogan21}, but we suggest that our results in this subsection should be interpreted with caution. A more robust approach would involve fitting the Freeman model, which is appropriate for thin rotating discs, to the asymmetric-drift corrected velocity field (i.e. $v_{\rm circ}$) and using this to derive the dynamical mass as a function of $R$. However, given the uncertainties on our beam-smearing corrections, the simplistic assumption that $\sigma_0$ holds constant across each galaxy, and the complex nature of our sources, such an approach is not justified and could lead to misleading results.

For sources in the KAOSS {\it disc-like} sample the median dynamical mass is $M_{\rm dyn}$\,$=$\,(1.1\,$\pm$\,0.2)\,$\times$\,10$^{11}$\,M$_\odot$, with a median stellar-to-dynamical mass ratio $M_\ast$/$M_{\rm dyn}$ of 1.4\,$\pm$\,0.3. Therefore our sources are consistent with having no dark matter within a central radius of $R$\,$\sim$\,7\,kpc. This is consistent with the suggestion that star-forming galaxies at high redshift are baryon dominated on scales of the disc \citep[e.g.,][]{genzel17,lang17}.

We therefore also consider the relation between dynamical mass and {\it total baryonic} mass, shown in Fig.~\ref{fig:dynamical_masses}. Baryonic masses are derived according to:
\begin{equation}
M_{\rm baryon} = M_\ast + \delta_{\rm gdr} M_{\rm dust},    
\end{equation}
where $\delta_{\rm gdr}$ is the gas-to-dust ratio for which we adopt a value of 65, using the fit of \cite{birkin21} to DSFGs at $z$\,$\sim$\,2. In contrast to the left panel of Fig.~\ref{fig:dynamical_masses}, we estimate dynamical masses within 4$R_{\rm e}$, rather than 2$R_{\rm e}$. From previous studies we expect $\sim$\,14\,kpc (the approximate median 4$R_{\rm e}$ of our sample) to encompass the majority of the molecular gas \citep[e.g.,][]{ivison11}. We demonstrate by the downwards arrow where the points would lie if we had adopted an aperture of 2$R_{\rm e}$ for the dynamical mass calculation. Although it is unclear how extended the cold molecular gas and dark matter is, Fig.~\ref{fig:dynamical_masses} suggests that much of it is situated beyond the stellar matter in DSFGs, and this is supported by our median $M_{\rm baryon}$/$M_{\rm dyn}$ ratio of 0.6\,$\pm$\,0.1, within 4$R_{\rm e}$.

We note here that changing the coefficient on the asymmetric drift term in Eq.~\ref{eq:v_circ} from 1 to 2 results in a $\sim$\,35 per cent increase in the dynamical mass at 2\,$R_{\rm e}$. This means that the dark matter fractions could vary from 0--35 per cent within $\sim$\,7\,kpc in the DSFGs, depending on the choice of this coefficient. Additionally, adopting the $v_{\rm circ}$ from {\sc galpak}$^{\rm 3D}$ for the nine galaxies we modelled produces an increase in the dynamical masses we derive (driven by the larger rotation velocities), as we find a median $M_{\rm dyn,GALPAK}/M_{\rm dyn}$\,$=$\,1.6\,$\pm$\,0.3 for these nine sources.

%
%
\subsection{Descendants of DSFGs}
\label{sec:descendants}

It has been suggested that DSFGs are connected with the progenitors of local massive and compact early-type galaxies \citep[e.g.,][]{lilly99,simpson14,toft14} in an evolutionary scenario involving obscured and unobscured QSO phases \citep[e.g.,][]{sanders88,blain02,swinbank06b, hopkins08}. Several authors have provided evidence supporting this claim, such as \citet{simpson14} who showed that $z$\,$\sim$\,0 DSFG descendants would have comparable stellar masses to massive early types (see also \citealt{dudzeviciute20}), and \cite{hodge16} found that ALESS DSFGs have gas surface densities and implied effective radii that are consistent with the most massive compact early-types. \cite{birkin21} compared CO-detected DSFGs with a sample of local early types in the Coma cluster \citep{shetty20} in the $M_{\rm baryon}$--$\sigma$ and Age--$\sigma$ plane, finding the two populations to be consistent. However, the $\sigma$ values used in that work were estimated statistically from sample-average CO linewidths without inclination corrections.\footnote{In this context, $\sigma$ is the \textit{effective} linewidth if all the kinetic energy of the galaxy was transferred from rotation-dominated to dispersion-dominated motion. In the remainder of this section we refer to $\sigma$ as $\sigma_{\rm eff}$.}

\begin{figure}
    \centering
    \includegraphics[width=\linewidth]{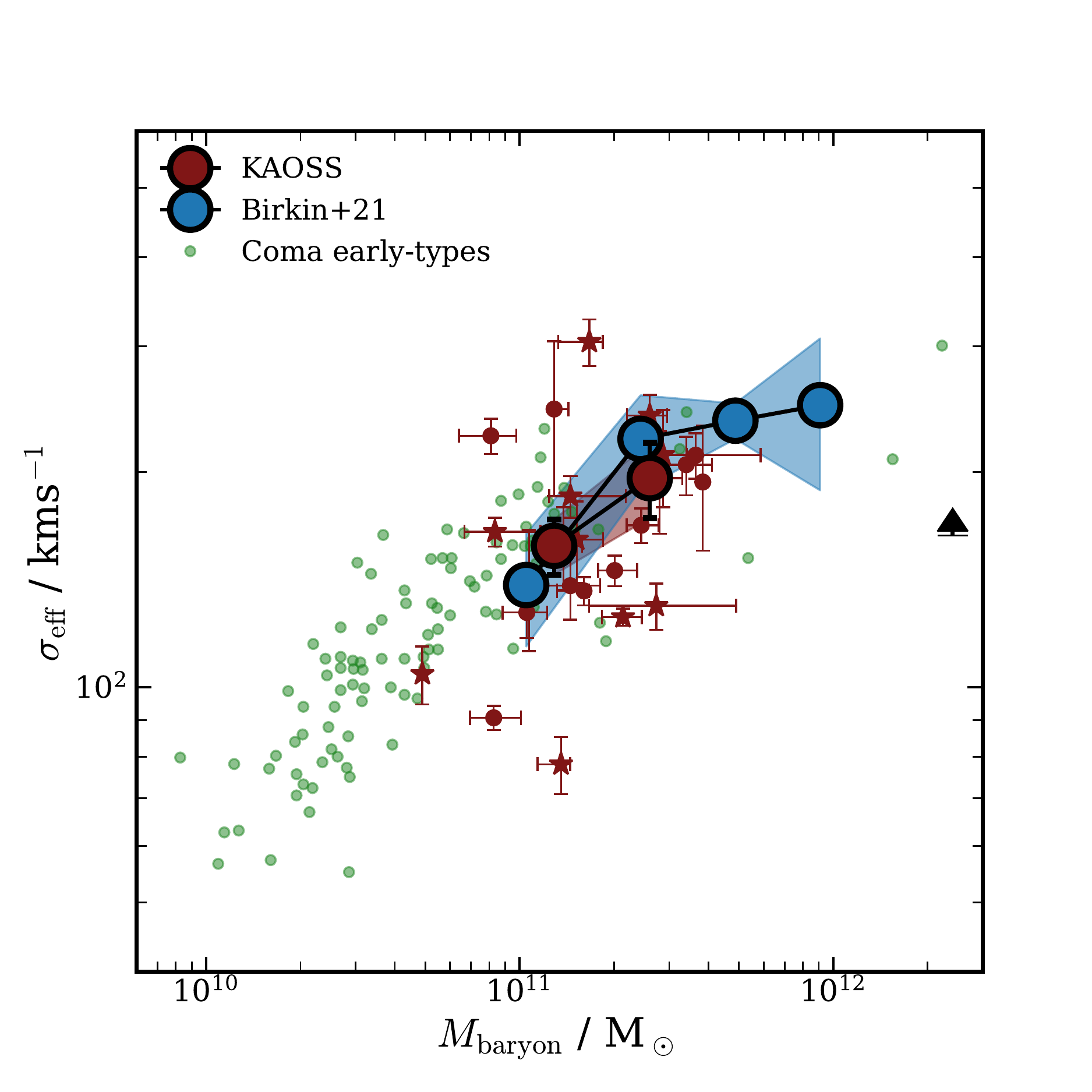}
    \caption{$M_{\rm baryon}$--$\sigma_{\rm eff}$ relation for KAOSS DSFGs (showing the individual points and their median trend by the large points and shaded region) and median trend of the CO-detected DSFGs from \protect\cite{birkin21}, and early-type galaxies from the Coma cluster \protect\citep{shetty20}. For KAOSS sources $M_{\rm baryon}$ is the sum of the {\sc magphys}-derived stellar and cold gas masses (see \S\ref{sec:mdyn}). Our sample appears to have comparable estimated velocity dispersions with the local early types, suggesting that they are plausibly connected with the descendant population of such sources.}
    \label{fig:mbaryon_sigma_kaoss}
\end{figure}

Our spatially resolved KMOS observations of DSFGs have provided inclination-corrected rotational and circular velocities (see \S\ref{sec:rot_curve_modelling}), from which we estimate $\sigma_{\rm eff}$ as \citep{galacticdynamics}:
\begin{equation}
    \sigma_{\rm eff} = \dfrac{v_{\rm circ}}{\sqrt{2}},
\end{equation}
to provide a more robust metric for comparing with the proposed low-redshift descendants. These values are presented in the online supplementary table. Fig.~\ref{fig:mbaryon_sigma_kaoss} shows $\sigma_{\rm eff}$ plotted against $M_{\rm baryon}$ for the KAOSS resolved subset, along with the trend observed in CO-detected sources from \cite{birkin21} and local early-type galaxies from \cite{shetty20}. We divide our sample into two bins in $M_{\rm baryon}$ and plot the median $\sigma_{\rm eff}$ with bootstrap uncertainties in these two bins, which are consistent with those of the CO sample, along with the early types, within their uncertainties. We see greater scatter among the KAOSS resolved sample compared to the local early type galaxies, however, this appears to be driven by AGN-hosting DSFGs. We conclude from Fig.~\ref{fig:mbaryon_sigma_kaoss} that our spatially resolved KMOS observations support the suggestion that $z$\,$\sim$\,2 DSFGs are consistent with being the progenitors of massive local early-type galaxies, which dominate the galaxy mass distribution at these masses \citep[e.g.,][]{ogle16}. We note that this result is robust against changing the coefficient on the asymmetric drift term in Eq.~\ref{eq:v_circ} from 1 to 2.

%
%
\section{Conclusions}
\label{sec:conclusions}

In this paper we have presented results from a subset of sources in the KMOS+ALMA Observations of Submillimetre Sources (KAOSS) Large Programme. We have measured spatially resolved H$\alpha$ velocity fields and extracted rotation curves for 27 ALMA-identified DSFGs in the COSMOS, UDS and GOODS-S fields, allowing us to derive rotational velocities and dynamical masses, along with $v_{\rm rot}/\sigma_0$ ratios, to test the level of rotational support in the DSFG population. Our main results are as follows:
\begin{itemize}
    \item We measure robust rotational velocities for a subsample of 24 out of the 27 resolved KAOSS sources from fitting Freeman disc models, finding a median inclination-corrected velocity at 2.2 times the disc radius of $v_{\rm rot}$\,$=$\,190\,$\pm$\,40\,km\,s$^{-1}$ and a median circular velocity $v_{\rm circ}$\,$=$\,230\,$\pm$\,20\,km\,s$^{-1}$. Therefore, 37\,$\pm$\,15 per cent of the DSFG sample are either not well described by disc-like kinematics, or have $v_{\rm rot}/\sigma_0$\,$<$\,1.5, and we make the assumption that these sources are pressure supported.
    \item We measure observed velocity dispersions, and by applying the beam-smearing corrections from \cite{johnson18}, we derive intrinsic velocity dispersions, $\sigma_0$. The KAOSS resolved sample has a median $\sigma_0$\,$=$\,87\,$\pm$\,6\,km\,s$^{-1}$, significantly higher than the samples of less actively star-forming galaxies at similar redshifts. This suggests high levels of turbulence in DSFGs.
    \item The median ratio of rotational velocity to intrinsic velocity dispersion is $v_{\rm rot}/\sigma_0$\,$=$\,2.5\,$\pm$\,0.5. This suggests that KAOSS DSFGs have significant rotational support but with a non-negligible contribution from pressure support.
    \item Our sources follow a trend between stellar mass $M_\ast$ and rotational velocity $v_{\rm rot}$ (the stellar Tully-Fisher relation), and we find a best-fit normalisation of the sTFR at $v_{\rm circ}$\,$=$\,100\,km\,s$^{-1}$ of $\log_{10}(M_\ast/{\rm M}_\odot)$\,$=$\,9.94\,$\pm$\,0.09, using a fixed slope of 3.4, which is consistent with the normalisation measured by \cite{tiley19} for $z$\,$=$\,0 galaxies, $\log_{10}(M_\ast/{\rm M}_\odot)$\,$=$\,9.87\,$\pm$\,0.04, at the same velocity.
    \item The KAOSS DSFGs have a median dynamical mass within 2$R_{\rm e}$ ($\sim$\,7\,kpc) of $M_{\rm dyn}$\,$=$\,(1.1\,$\pm$\,0.2)\,$\times$\,10$^{11}$\,M$_\odot$ and a median stellar-to-dynamical mass ratio of $M_\ast$/$M_{\rm dyn}$\,$=$\,1.4\,$\pm$\,0.3. Motivated by previous studies of the molecular gas in DSFGs, we estimate baryonic masses within a radius of 4$R_{\rm e}$ ($\sim$\,14\,kpc), finding a median baryonic-to-dynamical mass ratio of $M_{\rm bar}$/$M_{\rm dyn}$\,$=$\,0.6\,$\pm$\,0.1. We suggest that the majority of the baryonic matter in $z$\,$\sim$\,2 DSFGs is situated beyond the extent of the stellar emission.
    \item Using the inclination-corrected velocity dispersions we estimate effective stellar velocity dispersions for the KAOSS DSFGs, finding them to be consistent with early-type galaxies in the Coma cluster, and therefore potential progenitors of such sources.
\end{itemize}
The current resolved KAOSS sample has doubled the sample size of DSFGs with spatially resolved kinematics. With the completion of the survey, this sample should approximately double in size, giving us the ability to draw more statistical conclusions about the nature of the population. Current and upcoming instrumentation will enable further improvements. For example, the Enhanced Resolution Imager and Spectrograph (ERIS) on the VLT will enable adaptive-optics assisted IFU observations, providing kinematics via the H$\alpha$ emission line on $\sim$\,kpc scales. Similarly, with JWST/NIRSpec we will be able to detect the H$\alpha$ line and subsequently resolve DSFG kinematics out to higher redshifts, at a much higher resolution than that of the KAOSS data. This will allow us to search for multiple components which would provide direct evidence of ongoing mergers in DSFGs.

%
%
\section*{Acknowledgements}

We would like to thank Alfie Tiley for invaluable assistance with KMOS data reduction, along with Michael Hilker for technical support with our KMOS Large Programme. J.E.B. acknowledges the support of STFC studentship (ST/S50536/1). The Durham co-authors acknowledge support from STFC (ST/P000541/1, ST/T000244/1 and ST/X001075/1). C.-C.C. acknowledges support from the National Science and Technology Council of Taiwan (NSTC 109-2112-M-001-016-MY3 and 111-2112-M-001-045-MY3), as well as Academia Sinica through the Career Development Award (AS-CDA-112-M02). C.J.C. acknowledges support from the ERC Advanced Investigator Grant EPOCHS (788113). B.G. acknowledges support from the Carlsberg Foundation Research Grant CF20-0644 ‘Physical pRoperties of the InterStellar Medium in Luminous Infrared Galaxies at High redshifT: PRISMLIGHT. Y.M. acknowledges support of JSPS KAKENHI Grant Numbers JP17KK0098 and JP22H01273.

%
%
\section*{Data availability}

The data used in this paper are available through the ESO data archive. Reduced data products can be shared upon publication by reasonable request to the corresponding author.



\bibliographystyle{mnras}
\bibliography{bibliography}



\appendix

%
%
\section{Rotation curves}
\label{sec:rot_curves}

Fig.~\ref{fig:kaoss_rot_curves_1} shows the H$\alpha$ rotation curves and dispersion profiles of the 27 resolved KAOSS sources along with the derived velocity/velocity dispersion maps. These figures are discussed in more detail in \S\ref{sec:rot_curve_modelling}.

\begin{figure*}
    \centerline{\includegraphics[width=1\linewidth]{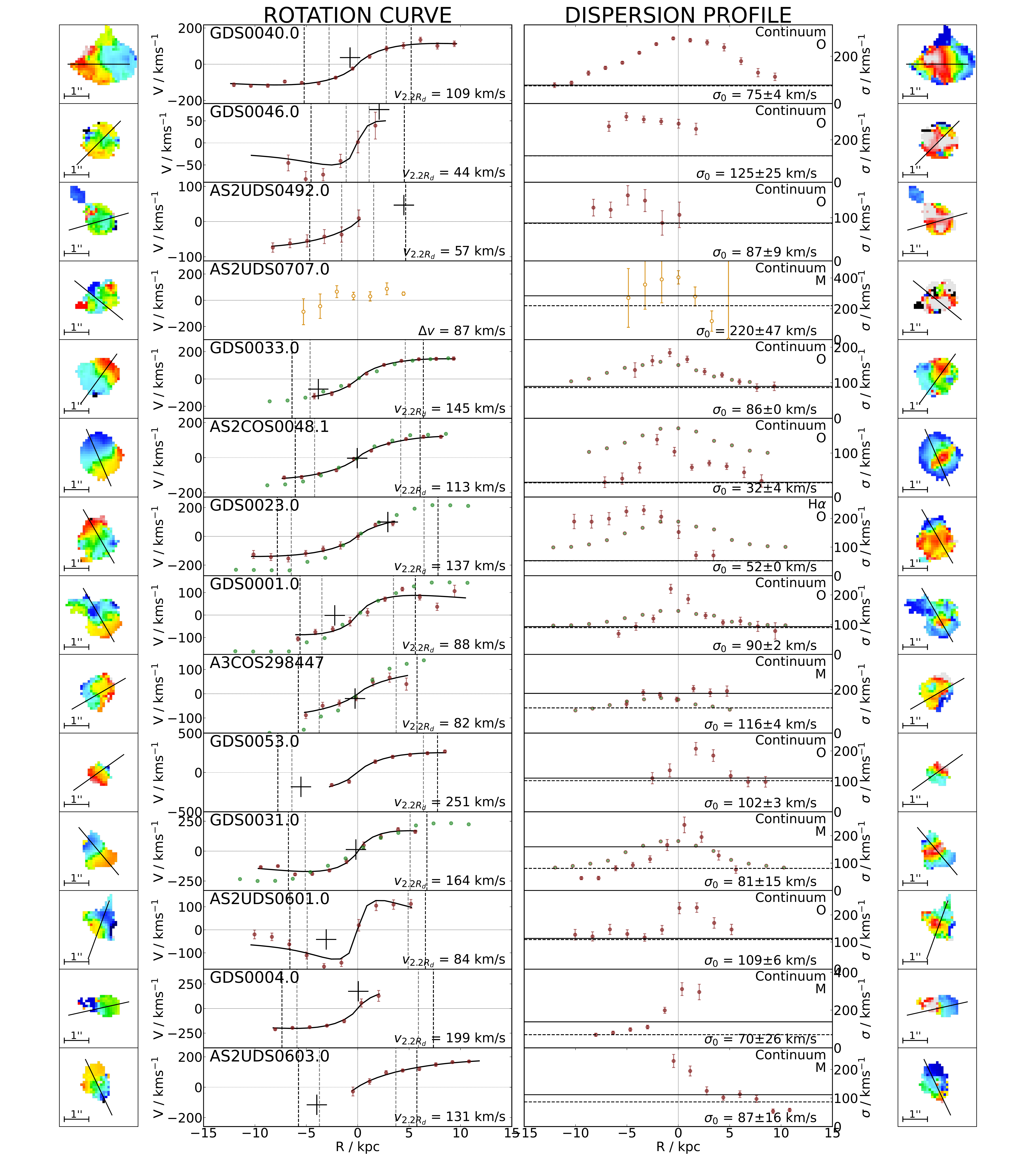}}
    \caption{H$\alpha$ rotation curves/maps (left) and velocity dispersion profiles/maps (right) for the 27 galaxies in the resolved KAOSS sample, shown with respect to the systemic velocity derived from the integrated H$\alpha$ spectra and the spatial centres derived in \S\ref{sec:rotation_axes}. Sources are ordered by the S/N of the H$\alpha$ emission line as in Fig.~\ref{fig:kaoss_vel_fields_1}. The solid black lines indicate Freeman disc model fits to the data, the vertical dashed lines indicate $\pm$\,2.2 times the disc radius (thin) and $\pm$\,2.2 times the disc radius convolved with $\sigma_{\rm PSF}$ (thick), the latter at which we measure the rotation velocity, and the cross indicates the position of the original kinematic and spatial centre before correction (i.e. $v_{\rm off}$ and $r_{\rm off}$). Points that are plotted as open circles are masked in the fitting procedure. For the dispersion profiles we indicate the observed velocity dispersion, $\sigma_{\rm obs}$ (horizontal black solid line), along with the method used to measure it (labelled in the top right of each panel; ``O'': outskirts or ``M'': median), and the beam smearing-corrected intrinsic velocity dispersion (horizontal black dashed line). We also indicate the method used to centre the velocity field in the top right corner of the dispersion profile panels. The green points show velocity/velocity dispersion profiles from {\sc galpak}$^{\rm 3D}$ (shown for nine sources).}
    \label{fig:kaoss_rot_curves_1}
\end{figure*}

\renewcommand{\thefigure}{A1 (Cont.)}
\begin{figure*}
    \ContinuedFloat
    \centerline{\includegraphics[width=1\linewidth]{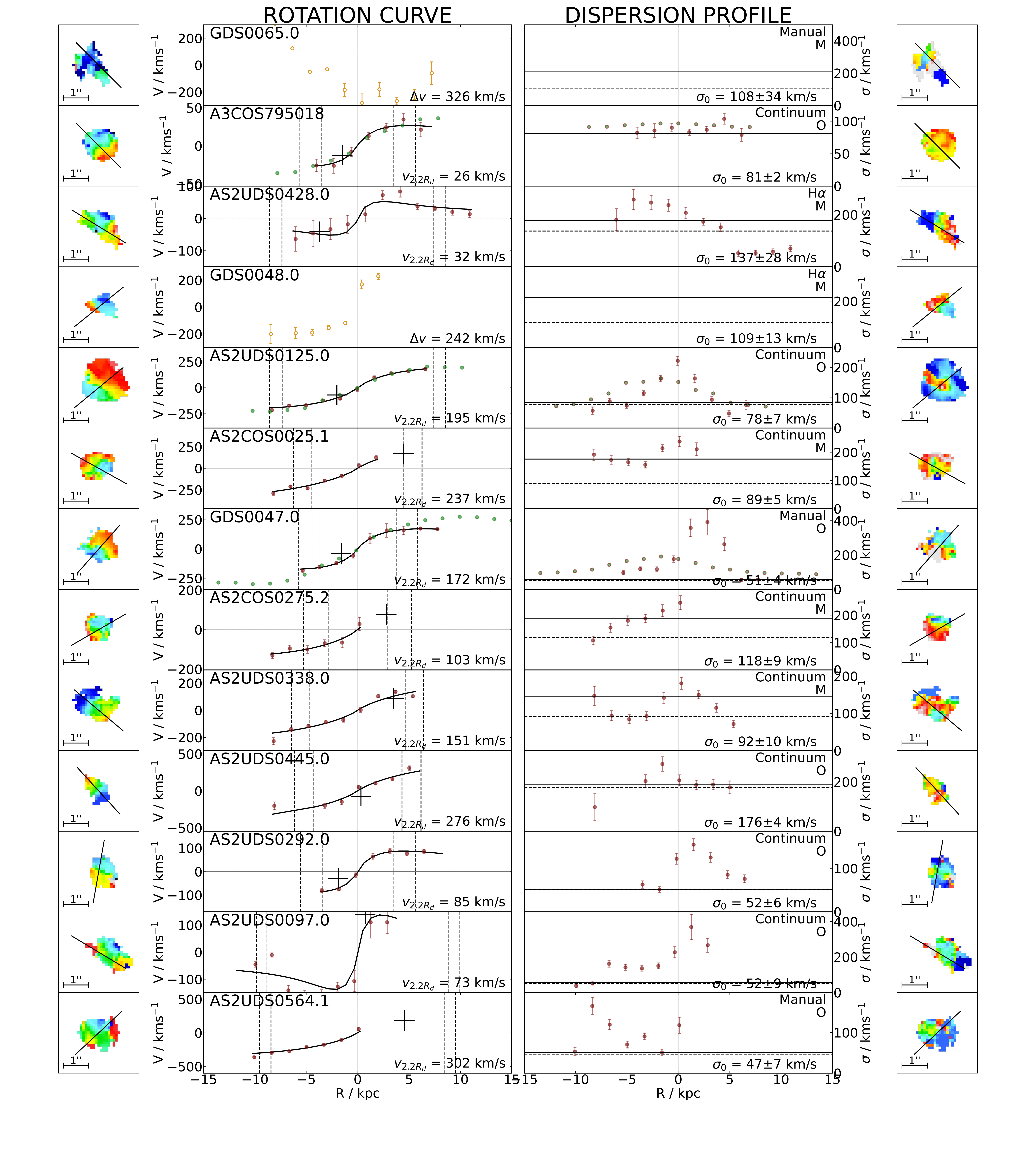}}
    \caption{}
    \label{fig:kaoss_rot_curves_2}
\end{figure*}
\renewcommand{\thefigure}{\arabic{figure}}

%
%
\section{Position-velocity diagrams}
\label{sec:rot_curves}

Fig.~\ref{fig:kaoss_pv_diagrams_major} shows H$\alpha$ major-axis position-velocity diagrams for all 27 KAOSS galaxies. A description of how these were derived is provided in \S\ref{sec:vel_maps}.

\renewcommand{\thefigure}{B1}
\begin{figure*}
    \centerline{\includegraphics[width=1\linewidth]{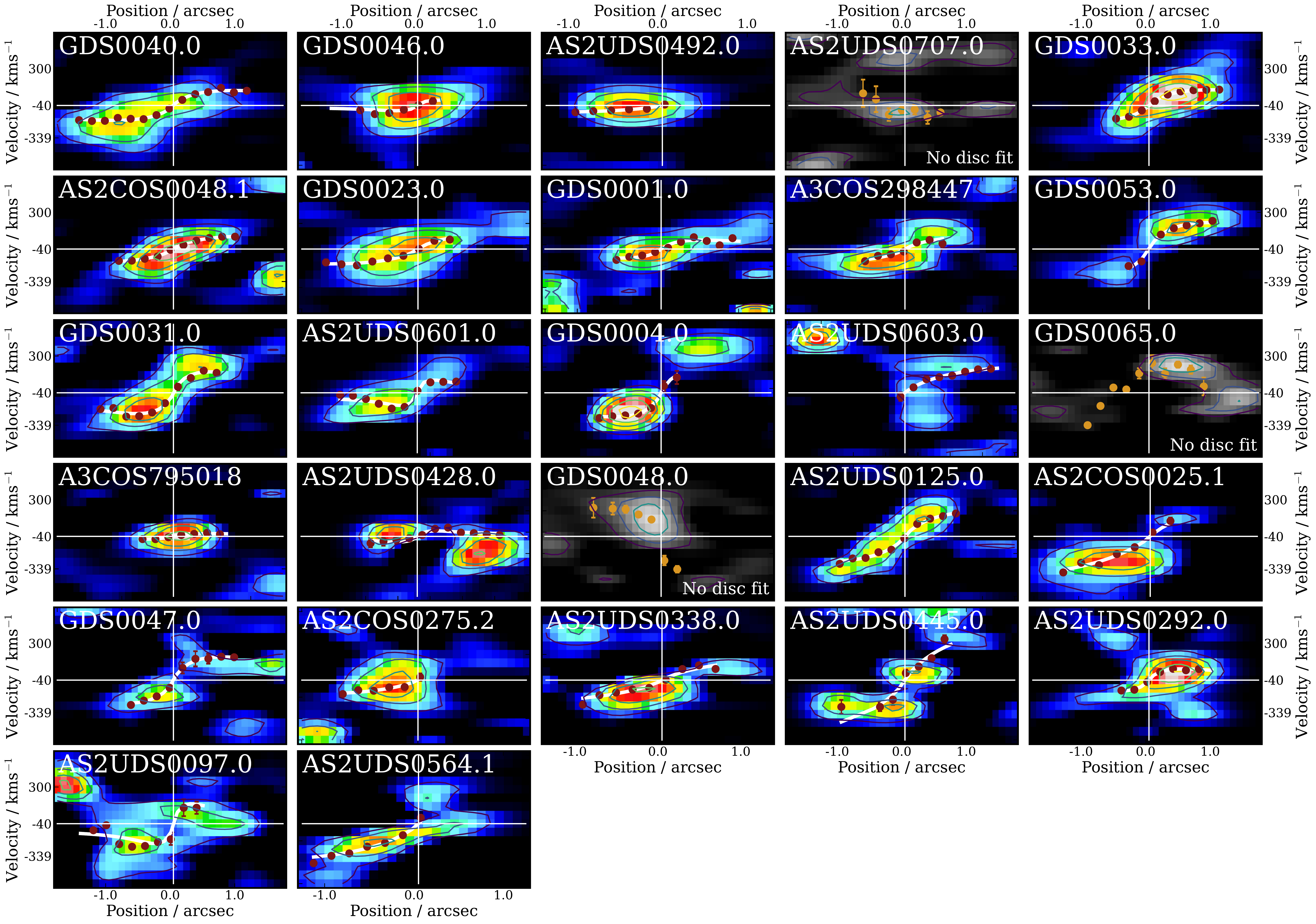}}
    \caption{Position-velocity (PV) diagrams for the 27 KAOSS galaxies. PV diagrams are extracted from a 0.5$''$ pseudo-slit along the kinematic major axis, and we smooth using a Gaussian window of FWHM corresponding to the approximate seeing in the $x$-direction (spatial) and velocity resolution in the $y$-direction (velocity). The colour scale shows the flux density of the emission, with the $x$-axis showing the position along the kinematic major axis and the $y$-axis showing the velocity offset from the systemic redshift. Three sources with rotation curves for which we were unable to fit a Freeman disc model are greyed out. We overlay the rotation curves derived in \S\ref{sec:rot_curve_modelling}, which generally trace the PV diagrams well, particularly in the higher S/N examples. For presentation purposes, the velocity axis labels are approximate, as the velocity scale varies with redshift.}
    \label{fig:kaoss_pv_diagrams_major}
\end{figure*}

%
%
\section{GALPAK$^{\rm 3D}$ models}
\label{sec:rot_curves}

Figs.~\ref{fig:galpak_fits_vel} and \ref{fig:galpak_fits_sig} show respectively the GALPAK$^{\rm 3D}$ best-fit velocity and velocity dispersions maps for the nine sources in the KAOSS sample which are modelled.

\renewcommand{\thefigure}{C1}
\begin{figure*}
    \centerline{\includegraphics[width=1\linewidth]{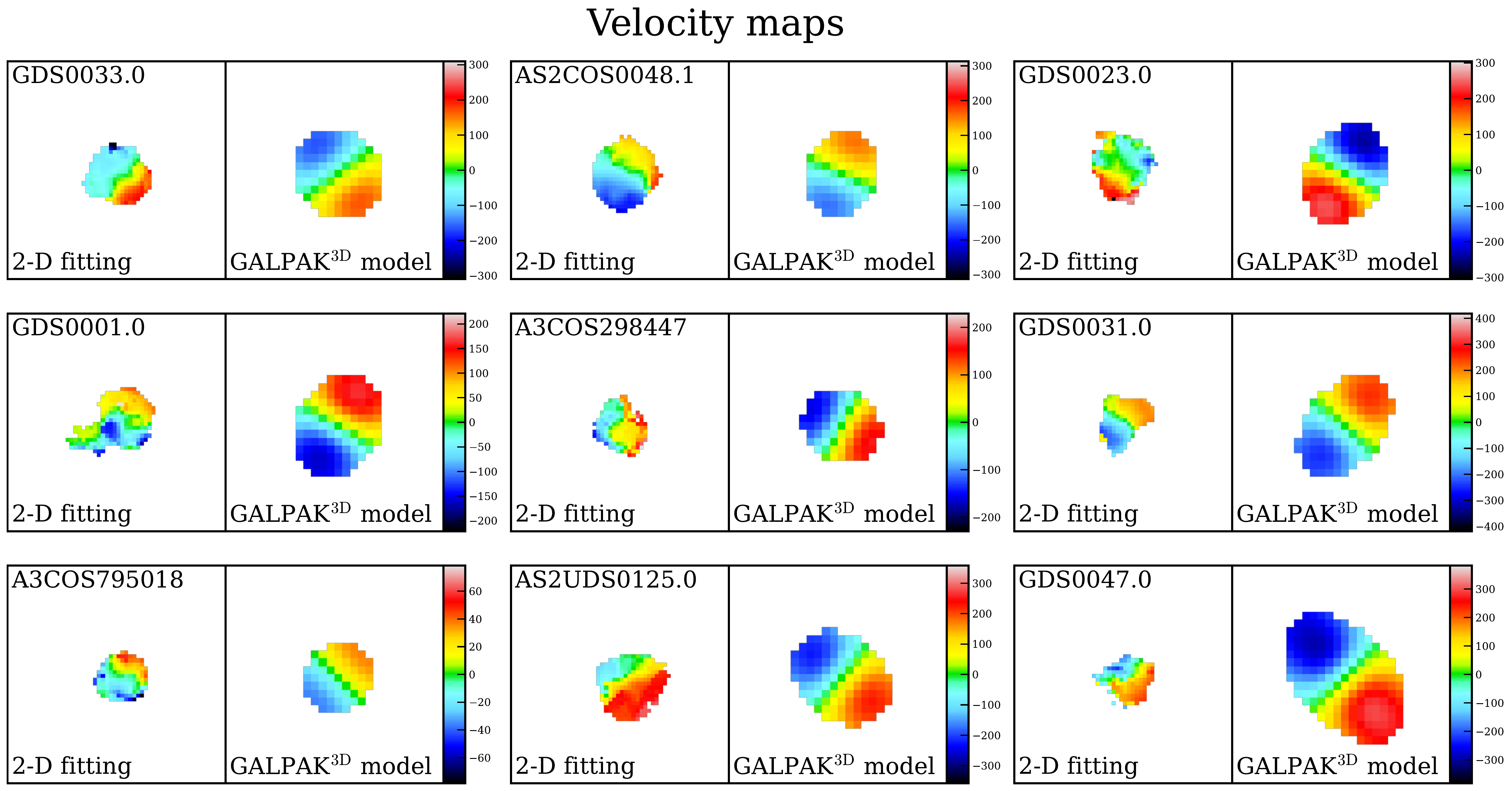}}
    \caption{Observed velocity maps for the nine sources that were modelled using GALPAK$^{\rm 3D}$ (uncorrected for beam smearing here), alongside the corresponding best-fit models. The color scale shows the velocity in km\,s$^{-1}$. We see that in the majority of cases GALPAK$^{\rm 3D}$ captures the structure of our empirically derived velocity maps, however the model fails to reproduce some of the more complex details, such as those seen in GDS0001.0 and GDS0023.0.}
    \label{fig:galpak_fits_vel}
\end{figure*}

\renewcommand{\thefigure}{C2}
\begin{figure*}
    \centerline{\includegraphics[width=1\linewidth]{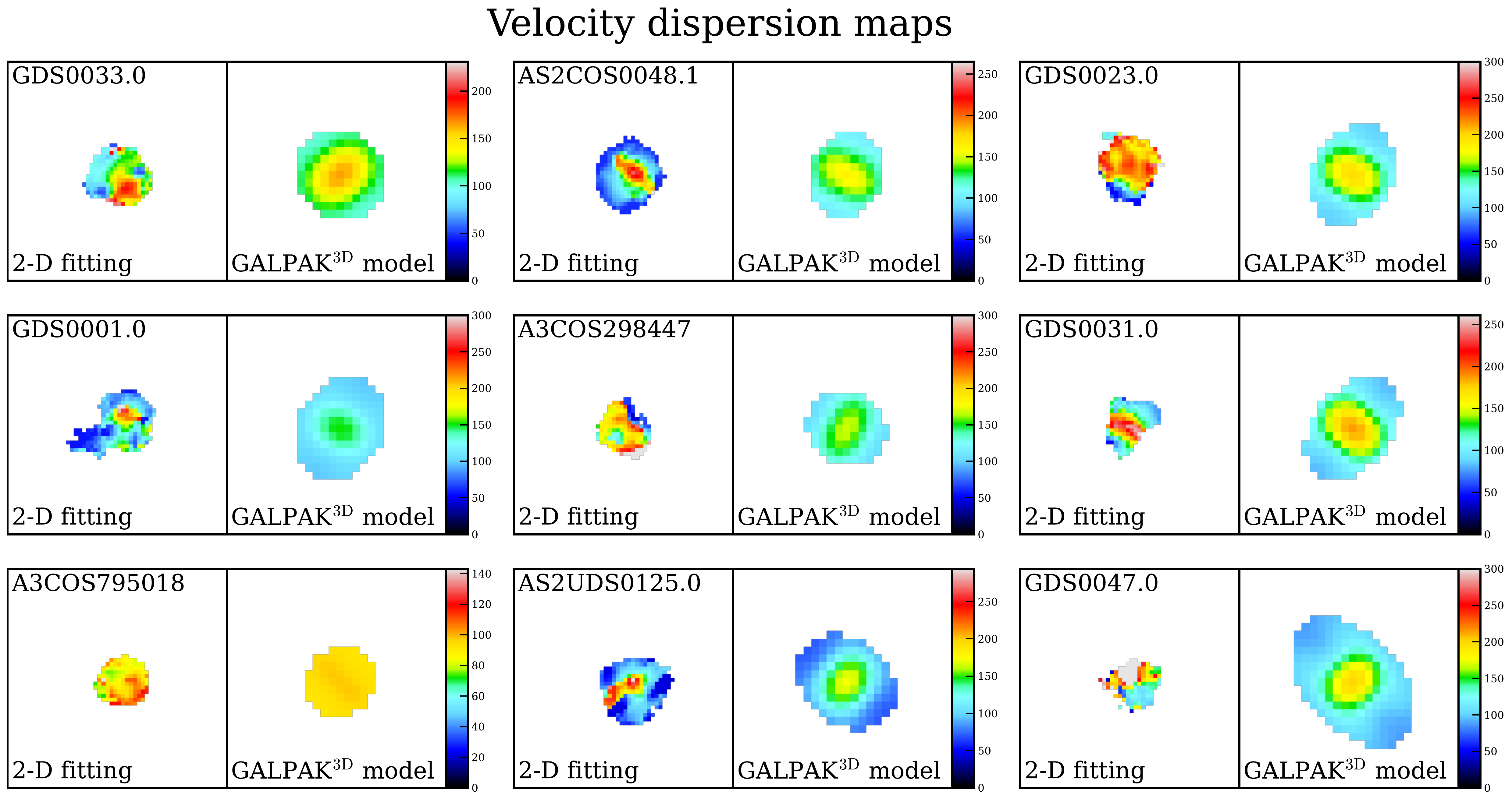}}
    \caption{Observed velocity dispersion fields for the nine sources that were modelled using GALPAK$^{\rm 3D}$ (uncorrected for beam smearing here), alongside the corresponding best-fit models. The color scale shows the velocity dispersion in km\,s$^{-1}$. We see that in all cases GALPAK$^{\rm 3D}$ fails to recover the structure of our empirically derived velocity dispersion maps, and we use this to suggest that the model is not suitable for application to these potentially complex systems and observations of modest resolution and S/N.}
    \label{fig:galpak_fits_sig}
\end{figure*}

\section*{Supporting information}
\label{sec:supporting_info}

Supplementary data are available at MNRAS online. We include a table of the properties used in this paper, along with a figure showing {\sc magphys} SED fits to the photometry of the KAOSS resolved sample.


\bsp	
\label{lastpage}

\end{document}